\documentclass{article}

\usepackage{arxiv}

\usepackage[utf8]{inputenc} 
\usepackage[T1]{fontenc}    
\usepackage{hyperref}       
\usepackage{url}            
\usepackage{booktabs}       
\usepackage{multirow}       
\usepackage{amsfonts}       
\usepackage{nicefrac}       
\usepackage{microtype}      
\usepackage{cleveref}       
\usepackage{graphicx}
\usepackage{natbib}
\usepackage{doi}
\usepackage{xcolor}
\usepackage{listings}
\usepackage{braket}         

\title{Qiskit QuantumKatas: Adapting Microsoft's Quantum Computing Exercises for LLM Evaluation}

\newif\ifuniqueAffiliation
\uniqueAffiliationtrue

\ifuniqueAffiliation
\author{
	Juan Cruz-Benito \\
	IBM Research \\
	\and
	Ismael Faro \\
	IBM Research \\
}
\fi

\renewcommand{\shorttitle}{Qiskit QuantumKatas for LLM Evaluation}

\hypersetup{
pdftitle={Qiskit QuantumKatas: Adapting Microsoft's Quantum Computing Exercises for LLM Evaluation},
pdfsubject={cs.CL, quant-ph, cs.SE},
pdfkeywords={Large Language Models, Quantum Computing, Qiskit, Benchmark, Code Generation},
}

\begin{document}
\maketitle

\begin{abstract}
We adapt Microsoft's QuantumKatas---a well-established quantum computing curriculum---from Q\# to Qiskit, the most widely-adopted quantum computing framework, and package it with an evaluation framework for systematic LLM assessment. The resulting benchmark comprises 350 tasks across 26 categories, spanning fundamental gates through advanced algorithms (Grover's, Simon's, Deutsch-Jozsa), error correction, key distribution, and quantum games. Each task includes a natural language prompt, canonical solution, and deterministic test verification via classical circuit simulation. By building on the QuantumKatas' proven pedagogical design rather than creating tasks from scratch, we inherit a principled difficulty progression and comprehensive concept coverage while contributing the framework adaptation, evaluation infrastructure, and empirical analysis.

We evaluate 16 LLMs across 7 prompting configurations---a total of 39,200 model runs---to demonstrate the benchmark's utility. Three key findings emerge: (1) the benchmark effectively differentiates model capabilities, with best-configuration pass rates ranging from 32.3\% to 83.1\% and a 26.1\,pp average gap between frontier and open-source models; (2) models perform well at implementing known algorithms (SimonsAlgorithm 82.1\%, BasicGates 81.6\%) but struggle with problem encoding (SolveSATWithGrover 34.4\%, DistinguishUnitaries 40.0\%); and (3) chain-of-thought prompting shows a modestly bimodal effect---it is the \textit{best} strategy for three models (two of them explicitly reasoning-tuned per vendor documentation) but degrades performance for the rest, leaving it mid-pack in aggregate (56.3\% mean) behind few-shot-5 (57.8\%). We release the benchmark, evaluation framework, and baseline results to support research on LLM capabilities in quantum computing.
\end{abstract}


\section{Introduction}

Large language models (LLMs) have demonstrated strong code generation capabilities across many programming languages and domains \citep{chen2021evaluating, austin2021program}, yet their performance on specialized scientific computing---particularly quantum computing---remains underexplored. Quantum computing poses a unique challenge: its fundamentally different computational paradigm requires understanding of superposition, entanglement, and measurement, concepts with no direct classical analogue.

Existing code generation benchmarks target general programming (HumanEval \citep{chen2021evaluating}, MBPP \citep{austin2021program}), data science (DS-1000 \citep{lai2023ds1000}), or software engineering (SWE-bench \citep{jimenez2024swebench}). Quantum-specific benchmarks have begun to emerge---Qiskit HumanEval \citep{qiskithumaneval2024} provides 150+ hand-curated tasks and QuantumBench \citep{quantumbench2024} offers multiple-choice questions---but there remains a need for larger-scale, pedagogically-structured benchmarks that enable fine-grained analysis of quantum programming capabilities.

We introduce the Qiskit QuantumKatas benchmark---a translation of Microsoft's QuantumKatas \citep{quantumkatas} from Q\# to Qiskit \citep{javadi2024quantum,gadi_aleksandrowicz_2019_2562111}, packaged with an evaluation framework for systematic LLM assessment. The original QuantumKatas are a well-established educational resource for learning quantum computing through hands-on programming, and we build directly on their design:

From the QuantumKatas we inherit 350 tasks across 26 categories (basic gates through Grover's search and quantum error correction) with a pedagogical progression that supports fine-grained capability assessment. Our contributions on top of this foundation are: (i)~a complete Qiskit translation \citep{unitaryfund2025survey} with an evaluation pipeline featuring deterministic verification via classical circuit simulation (statevector comparison), multi-provider LLM support, and configurable prompting strategies; (ii)~a large-scale empirical study of 16 models across 7 prompting configurations (39,200 runs), including prompting-strategy analysis and fine-grained profiling across 26 topics; and (iii)~analytical contributions: solution-diversity analysis via AST similarity, category-independence assessment, normalized difficulty metrics, and evidence that chain-of-thought prompting helps reasoning-tuned models but hurts others in this domain.

To validate the benchmark, we evaluate 16 LLMs across 7 prompting configurations. Best-configuration pass rates span 32.3\% to 83.1\%, and category-level analysis reveals that models implement known algorithms well (SimonsAlgorithm 82.1\%, BasicGates 81.6\%) but struggle with problem encoding (SolveSATWithGrover 34.4\%, DistinguishUnitaries 40.0\%). We also find that chain-of-thought prompting is modestly bimodal: it is the \textit{best} strategy for three models (two of them explicitly reasoning-tuned, GPT-5.3-Codex and Gemini 3.1 Pro, plus Gemma 4 26B-A4B) but degrades performance for the majority, leaving it mid-pack in aggregate (56.3\% mean) behind few-shot-5 (57.8\%).

\section{Related Work}

\subsection{Code Generation Benchmarks}

Several benchmarks have been developed to evaluate LLM code generation capabilities. HumanEval \citep{chen2021evaluating} introduced 164 Python programming problems with unit tests. MBPP \citep{austin2021program} expanded this with 974 crowd-sourced Python tasks. More recent benchmarks like DS-1000 \citep{lai2023ds1000} focus on data science libraries, while SWE-bench \citep{jimenez2024swebench} evaluates real-world software engineering tasks from GitHub issues.

\subsection{Scientific Computing and Reasoning Benchmarks}

Domain-specific benchmarks have also emerged: SciCode \citep{tian2024scicode} evaluates scientific programming across physics, chemistry, and biology, while GPQA \citep{gpqa2023} and MATH \citep{hendrycks2021math} test scientific and mathematical reasoning without code generation. CURIE \citep{cui2025curie} extends this multitask direction to long-context scientific reasoning across ten tasks spanning materials science, physics, quantum computing, and biology, with the best evaluated model reaching only 32\% accuracy. In the adjacent quantum-chemistry domain, QuantumChem-200K \citep{zeng2025quantumchem} provides a 200K-molecule dataset for fine-tuning LLMs on DFT-level property prediction, illustrating the quantum-flavored benchmarking effort outside quantum computing proper. Quantum computing---which requires \textit{both} scientific reasoning and specialized programming---has received comparatively limited attention despite its growing importance.

\subsection{Quantum Computing and LLMs}

IBM's Qiskit Code Assistant \citep{qiskitassistant2024,dupuis2024qiskit} represents an effort to create domain-specific LLMs for quantum computing. Several benchmarks have emerged for evaluating LLMs on quantum tasks, which we organize by what they measure.

\textit{Code generation benchmarks} evaluate the ability to produce working quantum programs. Qiskit HumanEval \citep{qiskithumaneval2024} introduced 150+ hand-curated Qiskit tasks with difficulty ratings, and Qiskit HumanEval-Hard \citep{qiskithumanevalhard2025} raised the bar by stripping import statements and boilerplate. QCoder \citep{qcoder2025} draws from quantum computing programming contest problems, emphasizing domain-specific metrics like circuit depth, while QuanBench \citep{quanbench2025} evaluates 44 Qiskit tasks using both functional correctness (Pass@K) and quantum semantic equivalence via process fidelity---finding that even the best models achieve below 40\% Pass@1. Multi-framework extensions such as QuanBench+ \citep{slim2026quanbenchplus} broaden this evaluation to aligned tasks across Qiskit, PennyLane, and Cirq, and PennyLang \citep{basit2025pennylang} releases a PennyLane-centric corpus of over 3{,}000 code samples paired with retrieval-augmented prompting.

\textit{Circuit and algorithm design benchmarks} focus on quantum circuit synthesis. QCircuitBench \citep{qcircuitbench2024} provides 120K+ data points across 25 algorithms for AI-driven circuit generation, and QHackBench \citep{qhackbench2025} uses PennyLane hackathon challenges for an alternative framework perspective. StabilizerBench \citep{paz2026stabilizerbench} narrows the focus to AI-assisted quantum error correction, providing tasks for stabilizer-circuit synthesis verifiable via the Gottesman--Knill formalism.

\textit{Conceptual understanding benchmarks} assess knowledge without requiring code. QuantumBench \citep{quantumbench2024} provides approximately 800 multiple-choice questions spanning nine areas of quantum science, and QC-Bench \citep{afane2026qcbench} extends this knowledge-evaluation direction to over 6{,}000 expert-level questions covering quantum algorithms, error correction, and security protocols, evaluated across 31 LLMs; the related Quantum-Audit preprint \citep{afane2026quantumaudit} additionally probes false-premise detection and topic-specific reasoning gaps across 26 models.

\textit{Adjacent quantum-LLM directions} target related modalities or skills. QCalEval \citep{cao2026qcaleval} benchmarks vision-language models on the interpretation of quantum-hardware calibration plots, and QuantumQA \citep{qu2026quantumqa} pairs a physics-consistent quantum-mechanics dataset with verification-aware reinforcement learning to improve LLM scientific reasoning. These efforts are complementary to code-generation benchmarks like ours, exercising different parts of the quantum-LLM stack.

Beyond benchmarking, recent work explores alternative paradigms. QUASAR \citep{quasar2025} applies agentic reinforcement learning with tool-augmented LLMs to OpenQASM 3.0 circuit generation, while M2QCode \citep{m2qcode2025} proposes a model-driven framework that generates quantum code for multiple platforms from UML-based models. In a parallel domain-specific direction, PennyCoder \citep{basit2025pennycoder} LoRA-fine-tunes a base LLM on PennyLane code (including QML and QRL) as an efficient on-device assistant analogous to Qiskit Code Assistant. On the data side, QuantumLLMInstruct \citep{kashani2024quantumllminstruct} releases a 500K-pair instruction-tuning corpus covering Hamiltonian construction, QASM generation, Jordan--Wigner mappings, and circuit decompositions.

Our benchmark complements these efforts with distinct characteristics: (1) pedagogical structure inherited from Microsoft's QuantumKatas enabling systematic difficulty progression, (2) fine-grained categorization (26 categories) for targeted capability analysis, (3) focus on Qiskit, the most widely-used quantum framework, and (4) comprehensive algorithm coverage from basic gates to Grover's, Simon's, and quantum error correction.

\section{The Qiskit QuantumKatas Benchmark}

\subsection{Dataset Construction}

The benchmark translates Microsoft's QuantumKatas \citep{quantumkatas}---an open-source, self-paced tutorial for learning quantum computing through Q\# programming exercises---into Qiskit, preserving the original pedagogical structure while adapting to Qiskit's API conventions.

\textbf{Choice of target framework.} We selected Qiskit for three reasons: (i)~it is the most widely-used quantum computing framework according to the Unitary Fund annual survey \citep{unitaryfund2025survey}; (ii)~Python-based frameworks are more commonly represented in LLM training data than Q\#, enabling fairer evaluation of general-purpose models; and (iii)~Qiskit's extensive documentation provides rich context that models are likely to have encountered during pre-training.

\textbf{Translation process.} AI coding agents (Claude Code \citep{claudecode2025} and Qiskit Code Assistant \citep{qiskitassistant2024,dupuis2024qiskit}) produced initial drafts of each translation. Every task was then manually reviewed by the authors for semantic faithfulness, Qiskit idiom correctness, and test adequacy. Approximately 30\% of tasks required non-trivial manual intervention, primarily in categories involving measurement semantics, ancilla qubit management, and multi-register circuit construction---areas in which Q\# and Qiskit conventions diverge most. The translation proceeded in four stages.

\begin{enumerate}
    \item \textbf{Task identification.} We extracted 350 distinct programming tasks from the QuantumKatas repository, covering 26 categories from basic gates to advanced algorithms.

    \item \textbf{API mapping.} Q\# operations were mapped to Qiskit equivalents:
    \begin{itemize}
        \item Q\#'s \texttt{X(q)} $\rightarrow$ Qiskit's \texttt{qc.x(q)}
        \item Q\#'s \texttt{CNOT(control, target)} $\rightarrow$ Qiskit's \texttt{qc.cx(control, target)}
        \item Q\#'s \texttt{Controlled X([controls], target)} $\rightarrow$ Qiskit's \texttt{qc.mcx(controls, target)}
        \item Q\#'s measurement and reset operations adapted to Qiskit's circuit model
    \end{itemize}

    \item \textbf{Test adaptation.} Q\#'s built-in assertion operations were translated to Python test functions using Qiskit's \texttt{Statevector} class and \texttt{AerSimulator} for verification. Each test validates correctness through statevector comparison or measurement-outcome analysis.

    \item \textbf{Validation.} All 350 canonical solutions were verified to pass their corresponding tests, ensuring that the translation preserved task semantics.
\end{enumerate}

\textbf{Task format.} Each translated task is a self-contained JSON record with four fields: a natural language prompt with mathematical notation (Unicode quantum kets, e.g., $|\psi\rangle$), a canonical solution, a test function with multiple test cases, and an entry point name. Type hints and docstrings follow Python conventions. \Cref{lst:task_format} shows an example.

\begin{lstlisting}[caption={Example task format (BasicGates/1.1)}, label={lst:task_format}]
{
  "task_id": "BasicGates/1.1",
  "prompt": "# Task: State flip\n# Input: A qubit in state |psi> = a|0> + b|1>\n# Goal: Change the state to a|1> + b|0>\n# Implement the function below:\ndef state_flip(qc, q):\n    pass",
  "canonical_solution": "def state_flip(qc, q):\n    qc.x(q)\n    return qc",
  "test": "def test_state_flip():\n    # Verification code using AerSimulator",
  "entry_point": "state_flip"
}
\end{lstlisting}

Tasks range from simple gate applications (e.g., BasicGates/1.1 ``State Flip,'' a single Pauli-X) to complex algorithm implementations (e.g., SolveSATWithGrover/3.1, which composes Boolean satisfiability encoding with Grover's search). Appendix~\ref{app:task_examples} shows one representative task per pedagogical tier.

\subsection{Task Categories}

The benchmark inherits Microsoft's pedagogical organization, spanning foundational concepts (BasicGates, Superposition, Measurements), canonical algorithms (Deutsch-Jozsa \citep{deutsch1992rapid}, Grover's search \citep{grover1996fast}, Simon's periodicity \citep{simon1997power}, QFT \citep{shor1999polynomial}), practical protocols (quantum teleportation \citep{bennett1993teleporting}, BB84 key distribution \citep{bennett2014quantum}, Superdense Coding), and advanced applications (quantum error correction \citep{shor1995scheme}, quantum games such as CHSH \citep{clauser1969proposed} and GHZ \citep{greenberger1989going}, oracle construction). \Cref{tab:categories} presents the full distribution.

\begin{table}[h]
\caption{Qiskit QuantumKatas benchmark task distribution across categories}
\label{tab:categories}
\centering
\begin{tabular}{lrl}
\toprule
Category & Tasks & Description \\
\midrule
BasicGates & 16 & Fundamental quantum gates (X, Y, Z, H, CNOT) \\
Superposition & 21 & Preparing superposition states \\
Measurements & 18 & Quantum measurements and outcomes \\
DeutschJozsa & 15 & Deutsch-Jozsa algorithm implementation \\
GroversAlgorithm & 8 & Grover's search algorithm \\
SimonsAlgorithm & 7 & Simon's periodicity algorithm \\
QFT & 16 & Quantum Fourier Transform \\
PhaseEstimation & 7 & Quantum Phase Estimation \\
RippleCarryAdder & 23 & Quantum arithmetic circuits \\
QEC\_BitFlipCode & 12 & Quantum error correction \\
KeyDistribution\_BB84 & 10 & BB84 quantum key distribution \\
Teleportation & 14 & Quantum teleportation protocols \\
DistinguishUnitaries & 15 & Distinguishing quantum operations \\
JointMeasurements & 13 & Multi-qubit measurements \\
MarkingOracles & 11 & Quantum oracle construction \\
GraphColoring & 17 & Graph coloring with quantum search \\
BoundedKnapsack & 17 & Knapsack problem encoding \\
CHSHGame & 8 & CHSH quantum game \\
GHZGame & 7 & GHZ quantum game \\
MagicSquareGame & 12 & Magic square game \\
SolveSATWithGrover & 10 & SAT solving with Grover's \\
SuperdenseCoding & 5 & Superdense coding protocol \\
TruthTables & 10 & Boolean function encoding \\
UnitaryPatterns & 18 & Unitary matrix patterns \\
tutorials & 32 & Introductory quantum concepts\textsuperscript{$\ast$} \\
examples & 8 & Worked example problems\textsuperscript{$\ast$} \\
\midrule
\textbf{Total} & \textbf{350} & \\
\bottomrule
\end{tabular}
\vspace{0.3em}
\par\noindent\footnotesize{\textsuperscript{$\ast$}These categories serve as pedagogical scaffolding rather than targeting specific quantum computing topics. We retain them because they contribute to the difficulty spectrum (tutorials: 82.2\% aggregate pass rate confirms calibration at the easy end) and because excluding them would misrepresent the QuantumKatas' scope. Results computed without these 40 tasks (310 tasks, 24 categories) do not materially change our findings.}
\end{table}

\subsection{Dataset Statistics}

\Cref{tab:statistics} summarizes key dataset characteristics that inform task complexity and coverage.

\begin{table}[h]
\caption{Dataset statistics characterizing task complexity and coverage}
\label{tab:statistics}
\centering
\begin{tabular}{lr}
\toprule
Metric & Value \\
\midrule
\multicolumn{2}{l}{\textit{Scale}} \\
Total tasks & 350 \\
Categories & 26 \\
Avg tasks per category & 13.5 \\
\midrule
\multicolumn{2}{l}{\textit{Solution complexity (lines of code)}} \\
Minimum & 5 \\
Maximum & 148 \\
Average & 31.2 \\
\midrule
\multicolumn{2}{l}{\textit{Pedagogical difficulty (curriculum-based)}} \\
Introductory & 95 (27.1\%) \\
Intermediate & 132 (37.7\%) \\
Advanced & 123 (35.1\%) \\
\bottomrule
\end{tabular}
\end{table}

\textbf{Difficulty distribution.} Rather than relying on lines of code as a proxy for difficulty, we classify tasks into three tiers. The QuantumKatas were designed as a structured learning path but do not include explicit difficulty labels; our tiering follows the order in which Microsoft introduces categories in the original curriculum (foundational topics first, canonical algorithms next, compositional/encoding tasks last) and respects their conceptual prerequisites. The three tiers are as follows.

\begin{itemize}
    \item \textbf{Introductory} (95 tasks, 27.1\%). Entry points to quantum computing---tutorials, worked examples, BasicGates, Superposition, and Measurements---covering single-qubit operations, basic state preparation, and foundational measurement concepts.

    \item \textbf{Intermediate} (132 tasks, 37.7\%). Canonical quantum algorithms and protocols that build on foundational concepts: Deutsch--Jozsa, Simon's, Grover's, QFT, phase estimation, teleportation, superdense coding, quantum key distribution (BB84), quantum error correction (bit-flip code), joint (multi-qubit) measurements, quantum games (CHSH, GHZ), and Boolean function encoding.

    \item \textbf{Advanced} (123 tasks, 35.1\%). Categories requiring composition of multiple quantum concepts or problem encoding: quantum arithmetic (RippleCarryAdder), unitary discrimination (DistinguishUnitaries), oracle construction (MarkingOracles), constraint satisfaction (GraphColoring, BoundedKnapsack, SolveSATWithGrover), quantum games requiring entanglement strategies (MagicSquareGame), and complex unitary synthesis (UnitaryPatterns).
\end{itemize}

This classification reflects the conceptual prerequisites and compositional complexity of each category rather than the length of its solution code. The evaluation results are consistent with this ordering: introductory categories average a 65.7\% pass rate, intermediate categories 61.9\%, and advanced categories 50.9\% (\Cref{fig:pedagogical_tiers}), confirming that the pedagogical progression translates into measurable difficulty for LLMs.

\textbf{Concept coverage.} Whole-word, case-insensitive keyword matching over task prompts gives: oracles (18.0\% of tasks; keyword \texttt{oracle}), superposition (12.3\%; \texttt{superposition}), measurement (11.7\%; \texttt{measurement}), phase manipulation (11.7\%; \texttt{phas*}), unitary operations (5.7\%; \texttt{unitary}), and controlled operations (5.1\%; \texttt{controlled}). The exact keywords are listed so readers can reproduce the counts from the dataset.

\subsection{Comparison to Existing Benchmarks}

\Cref{tab:benchmarks} positions the Qiskit QuantumKatas benchmark relative to existing benchmarks.

\begin{table}[h]
\caption{Comparison with existing benchmarks for LLM evaluation}
\label{tab:benchmarks}
\centering
\small
\begin{tabular}{lcccl}
\toprule
Benchmark & Tasks & Domain & Verification & Focus \\
\midrule
HumanEval \citep{chen2021evaluating} & 164 & General Python & Unit tests & Algorithmic \\
MBPP \citep{austin2021program} & 974 & General Python & Unit tests & Basic coding \\
DS-1000 \citep{lai2023ds1000} & 1,000 & Data science & Execution & Library usage \\
SWE-bench \citep{jimenez2024swebench} & 2,294 & Software eng. & Test suite & Bug fixing \\
SciCode \citep{tian2024scicode} & 338 & Scientific & Numerical & Multi-domain \\
CURIE \citep{cui2025curie} & 580 & Scientific & Numerical & Multitask reasoning \\
\midrule
Qiskit HumanEval \citep{qiskithumaneval2024} & 150+ & Quantum & Simulation & QC tasks \\
QuantumBench \citep{quantumbench2024} & 800 & Quantum & MCQ & QC concepts \\
QC-Bench \citep{afane2026qcbench} & 6{,}000+ & Quantum & MCQ & QC knowledge \\
QCircuitBench \citep{qcircuitbench2024} & 120K+ & Quantum & Simulation & Algorithm design \\
QHackBench \citep{qhackbench2025} & --- & Quantum & Simulation & PennyLane tasks \\
QCoder \citep{qcoder2025} & 58 & Quantum & Simulation & Contest problems \\
QuanBench \citep{quanbench2025} & 44 & Quantum & Sim.+Fidelity & Algorithms \\
QuanBench+ \citep{slim2026quanbenchplus} & 42 & Quantum & Simulation & Cross-framework \\
\midrule
\textbf{Qiskit QuantumKatas} & \textbf{350} & \textbf{Quantum} & \textbf{Simulation} & \textbf{QC curriculum} \\
\bottomrule
\end{tabular}
\vspace{0.5em}
\par\noindent\footnotesize{Note: ``---'' indicates task count not specified in the source publication (QHackBench aggregates problems from hackathon challenges).}
\end{table}

The benchmark builds on three properties inherited from Microsoft's QuantumKatas---pedagogical structure (curriculum from basic gates to advanced algorithms), category granularity (26 categories for fine-grained analysis), and comprehensive algorithm coverage (Deutsch-Jozsa, Grover's, Simon's, error correction, quantum games). Our contributions layer on top of this foundation: a Qiskit translation making this curriculum accessible on the most widely-used framework (350 tasks, approximately 2$\times$ Qiskit HumanEval), a complete evaluation pipeline with deterministic verification, and baseline results from 16 models across 7 configurations.

\textbf{Use cases and scope.} The benchmark supports LLM evaluation with 26-category granularity, model development (canonical solutions as supervised targets), prompting research, quantum-education tooling, and cross-framework adaptation. It evaluates code generation via classical simulation, not real quantum hardware, so it does not assess noise mitigation, decoherence handling, or hardware-specific optimization. We also flag a dual-use concern: widespread LLM access to kata solutions may undermine their pedagogical value---educators should pair automated feedback with assessments requiring conceptual explanation, not just working code.

\section{Evaluation Framework}

\subsection{Methodology}

Our evaluation framework supports multiple LLM providers through a unified interface. For each task, we:

\begin{enumerate}
    \item Present the task prompt to the model with the following system prompt:
    \begin{quote}
    \textit{``You are an expert quantum computing programmer specializing in Qiskit. Your task is to implement quantum computing functions using Qiskit. Provide ONLY the Python code implementation, no explanations. The code should be complete and ready to execute.''}
    \end{quote}

    \item Extract Python code from the model's response using a multi-stage parser that first attempts markdown code blocks (\verb|```python ... ```|), then triple-quoted strings, and falls back to the raw response.

    \item Validate syntax using Python's \texttt{ast.parse()} to catch syntax errors before execution.

    \item Execute the solution in an isolated subprocess with a 30-second timeout, captured stdout, and a full Qiskit environment with \texttt{AerSimulator}. This timeout accommodates all canonical solutions (the longest of which completes in under 10 seconds); only 7 of 39,200 evaluation runs (0.018\%) exceeded this limit.

    \item Verify correctness by running the task's test function, which constructs test circuits, executes them through classical simulation (\texttt{AerSimulator}), and compares output states against expected values via statevector or measurement verification.
\end{enumerate}

All API calls include retry logic with exponential backoff for rate limits and transient errors.

\subsection{Models Evaluated}

We evaluate 16 distinct models spanning two categories.

\textbf{Frontier Models (Proprietary):}
\begin{itemize}
    \item \textbf{Claude family} \citep{anthropic2024claude}: Claude Opus 4.7, Claude Sonnet 4.6, Claude Haiku 4.5
    \item \textbf{GPT family} \citep{openai2025gpt5}: GPT-5.5, GPT-5.3-Codex
    \item \textbf{Gemini family}: Gemini 3.1 Pro \citep{gemini3pro2025}
\end{itemize}

\textbf{Open-Source Models.} We group by total parameter count, with mixture-of-experts (MoE) active counts noted where applicable; the cohort is bimodal, with no model in the 100--400B range, so a two-tier split is the most informative grouping.
\begin{itemize}
    \item \textbf{Large models} ($\geq$100B total): Mistral Large 3 (675B) \citep{mistrallarge2025}, Llama 4 Maverick (400B total, 17B active) \citep{llama4_2025}, GPT-OSS-120B \citep{gptoss2025}, Llama 4 Scout (109B total, 17B active) \citep{llama4_2025}
    \item \textbf{Small models} ($<$100B total): Gemma 4 31B \citep{gemma4_2026}, Granite 4.1 30B \citep{granite41_2026}, Gemma 4 26B-A4B (26B total, 4B active) \citep{gemma4_2026}, Mistral Small 3.2 24B \citep{mistralsmall2025}, GPT-OSS-20B \citep{gptoss2025}, Granite 4.1 8B \citep{granite41_2026}
\end{itemize}

All models were evaluated with temperature 0 (or 1.0 for reasoning models that require it, per provider documentation) for reproducibility. We note that temperature 0 does not guarantee fully deterministic outputs across all providers---implementation details such as batching and floating-point non-determinism can introduce minor run-to-run variation. We performed single-run evaluations rather than multiple trials, so the confidence intervals reported in \Cref{tab:results} reflect binomial uncertainty over the task population, not run-to-run variance.

We report pass@1 (single-attempt) results throughout this paper. Standard code generation benchmarks often report pass@k for $k > 1$ at non-zero temperature \citep{chen2021evaluating}, which captures a model's ability to produce \textit{at least one} correct solution across multiple samples---a measure that can diverge substantially from pass@1. Our use of temperature 0 makes repeated sampling near-identical, so pass@k $\approx$ pass@1 under our setup. However, this means our results represent a \textit{lower bound} on model capability: models that narrowly fail on some tasks at temperature 0 might succeed with diverse sampling. Evaluating pass@5 at non-zero temperature for a representative subset of models is a concrete direction for future work that would improve comparability with the broader code generation literature and better characterize the gap between deterministic and stochastic evaluation on quantum tasks.

\textbf{Compute budget.} The full evaluation comprised 39,200 API calls (350 tasks $\times$ 16 models $\times$ 7 configurations) plus local test execution for each response. Total wall-clock time was approximately two weeks, dominated by API rate limits and sequential per-model execution rather than local computation. We estimate the aggregate API billing cost at roughly \$500--\$700 USD for the commercially hosted models (Anthropic, OpenAI/Azure, and Google), though exact figures vary by provider pricing and token consumption; the remaining models were served on internal infrastructure with no per-token charges.

\subsection{Prompting Strategies}

We evaluate each model across 7 prompting configurations:

\begin{itemize}
    \item \textbf{Zero-shot}: Three system prompt variants (default, minimal, detailed)
    \item \textbf{Few-shot}: 1-shot, 3-shot, and 5-shot with solved examples drawn from introductory categories (BasicGates and Superposition). Examples were selected deterministically by iterating through the dataset and taking the first $k$ tasks from these categories whose canonical solutions had been verified, ensuring reproducibility across runs. The same examples were used for all models and all target categories. The current task being evaluated is always excluded from the example set, preventing direct leakage.
    \item \textbf{Chain-of-thought}: Explicit reasoning steps before code generation
\end{itemize}

\textbf{Few-shot design rationale.} We chose to draw examples from introductory categories rather than from the same category being tested. This design avoids information leakage---providing a Simon's algorithm example before testing another Simon's task would reveal algorithmic structure---while still demonstrating Qiskit coding patterns and function signature conventions. The tradeoff is that examples may be less relevant to advanced tasks; same-category examples might yield higher few-shot gains, but would conflate algorithmic hint-giving with genuine few-shot learning. We acknowledge that few-shot performance can be sensitive to example selection \citep{wei2022chain}. Our cross-category design is deliberately conservative: same-category examples would likely yield higher few-shot gains (by providing algorithmic hints), meaning the modest improvements we report (+2.4\,pp from zero-shot default to few-shot-5 on average) are plausibly a lower bound on what few-shot prompting can achieve for quantum code generation. Exploring alternative strategies (e.g., same-category, difficulty-matched, or diversity-maximizing examples) and quantifying this gap is an important direction for future work.

\section{Results}

We report results along four axes: overall model performance, the effect of prompting strategy, category-level difficulty profiling, and a taxonomy of failure modes. Best-configuration pass rates span 32.3\% to 83.1\% across the 16 models, with frontier models averaging 26.1\,pp above open-source; few-shot-5 is the most reliable strategy in aggregate (57.8\% mean) while chain-of-thought exhibits a modestly bimodal per-model effect; and category pass rates span 34.4\% (SolveSATWithGrover) to 85.4\% (UnitaryPatterns), with algorithm implementation systematically outperforming problem encoding.

\subsection{Overall Performance}

\Cref{tab:results} presents the benchmark results for all 16 models, showing best configuration performance with 95\% Wilson score confidence intervals \citep{wilson1927probable}.

\begin{table}[h]
\caption{Overall benchmark results ranked by best configuration. 95\% Wilson score confidence intervals in brackets. Avg shows mean across all 7 configurations. Frontier models marked with $\dagger$; others are open-source. For Gemma 4 31B and Granite 4.1 8B, few-shot-1 and few-shot-5 tie at the reported pass rate; we list few-shot-5 in the Best Config column but either is a valid best.}
\label{tab:results}
\centering
\small
\begin{tabular}{rlccc}
\toprule
Rank & Model & Best [95\% CI] & Avg & Best Config \\
\midrule
1 & GPT-5.5$^\dagger$ & \textbf{83.1\%} [78.9, 86.7] & 80.8\% & zero-shot (default) \\
2 & Claude Opus 4.7$^\dagger$ & 80.9\% [76.4, 84.6] & 79.4\% & few-shot-3 \\
3 & Claude Sonnet 4.6$^\dagger$ & 78.0\% [73.4, 82.0] & 73.3\% & few-shot-5 \\
4 & GPT-5.3-Codex$^\dagger$ & 75.1\% [70.4, 79.4] & 70.8\% & chain-of-thought \\
5 & Gemini 3.1 Pro$^\dagger$ & 74.6\% [69.8, 78.8] & 68.8\% & chain-of-thought \\
6 & Gemma 4 31B (31B) & 68.0\% [62.9, 72.7] & 66.1\% & few-shot-5 \\
7 & GPT-OSS-120B (117B / 5B active) & 65.7\% [60.6, 70.5] & 61.8\% & zero-shot (default) \\
8 & Gemma 4 26B-A4B (26B / 4B active) & 61.4\% [56.2, 66.4] & 58.2\% & chain-of-thought \\
9 & Claude Haiku 4.5$^\dagger$ & 60.3\% [55.1, 65.3] & 58.2\% & few-shot-1 \\
10 & GPT-OSS-20B (21B / 3.6B active) & 59.7\% [54.5, 64.7] & 53.9\% & few-shot-3 \\
11 & Mistral Large 3 (675B) & 48.6\% [43.4, 53.8] & 45.1\% & few-shot-5 \\
12 & Llama 4 Maverick (400B / 17B active) & 43.7\% [38.6, 49.0] & 39.3\% & few-shot-5 \\
13 & Mistral Small 3.2 24B (24B) & 40.0\% [35.0, 45.2] & 36.1\% & few-shot-3 \\
14 & Granite 4.1 30B (30B) & 38.9\% [33.9, 44.1] & 36.5\% & few-shot-5 \\
15 & Llama 4 Scout (109B / 17B active) & 34.3\% [29.5, 39.4] & 30.3\% & few-shot-5 \\
16 & Granite 4.1 8B (8B) & 32.3\% [27.6, 37.4] & 28.8\% & few-shot-5 \\
\bottomrule
\end{tabular}
\end{table}

\begin{figure}[h]
\centering
\includegraphics[width=\textwidth]{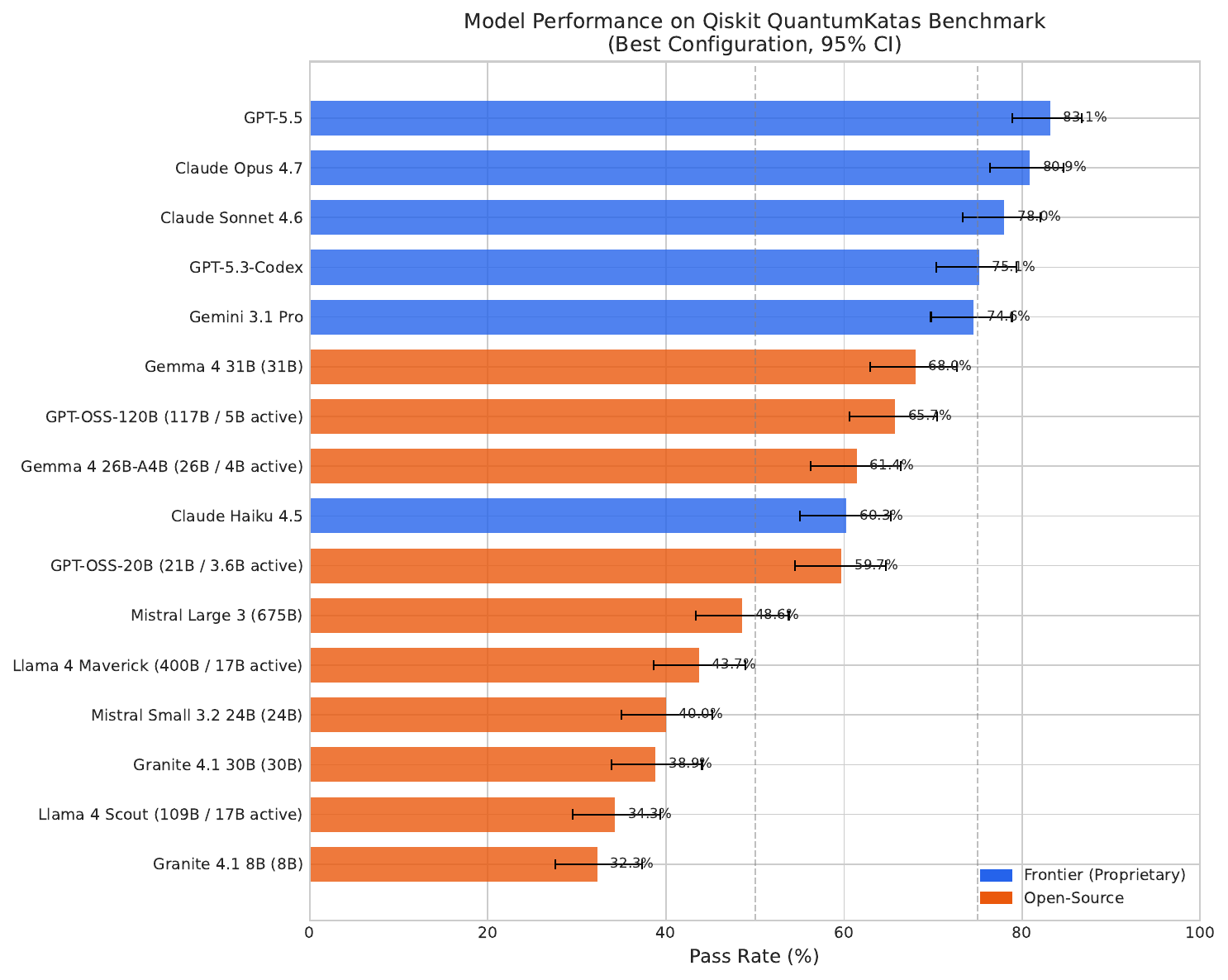}
\caption{Model performance on the Qiskit QuantumKatas benchmark. Bars show pass rates for each model's best configuration, with error bars indicating 95\% Wilson score confidence intervals. Blue bars indicate frontier (proprietary) models; orange bars indicate open-source models.}
\label{fig:model_rankings}
\end{figure}

\Cref{fig:model_rankings} visualizes these results, showing pass rates with 95\% confidence intervals for all models, distinguishing frontier (proprietary) from open-source models.

Several patterns emerge from \Cref{tab:results} and \Cref{fig:model_rankings}.

\begin{itemize}
    \item \textbf{Frontier models dominate the top.} GPT-5.5 achieves the highest point estimate (83.1\%, 95\% CI: 78.9--86.7\%), followed by Claude Opus 4.7 (80.9\%) and Claude Sonnet 4.6 (78.0\%); GPT-5.5 and Claude Opus 4.7 are not statistically distinguishable at the 95\% level. The top tier ($>$70\%) is populated exclusively by frontier models, and no open-source model reaches that threshold: Gemma 4 31B (68.0\%) is the strongest open-source entry.

    \item \textbf{Scale does not guarantee performance.} Size is a weak predictor within the open-source tier. Mistral Large 3 (675B parameters, 48.6\%) is outperformed by Gemma 4 31B (68.0\%) and GPT-OSS-120B (65.7\%), both roughly one-tenth its size. Training recipe and data composition appear to matter more than raw parameter count for quantum programming.

    \item \textbf{Model family consistency.} The GPT family is notably robust across configurations (GPT-5.5: 83.1\% best vs 80.8\% avg, $\Delta$=2.3\,pp; GPT-OSS-120B: $\Delta$=4.0\,pp), as is Claude Opus 4.7 ($\Delta$=1.4\,pp). All 16 models in the cohort cluster within roughly 6\,pp of best-vs-average configuration delta---the largest gaps are GPT-OSS-20B and Gemini 3.1 Pro at $\Delta$=5.8\,pp---indicating that no model in this cohort relies on a single ``magic'' prompt to reach its reported pass rate.

    \item \textbf{Reasoning-tuned models prefer CoT.} Three of the 16 models achieve their best configuration under chain-of-thought prompting: GPT-5.3-Codex, Gemini 3.1 Pro, and Gemma 4 26B-A4B. Two of these are explicitly reasoning-oriented model variants per vendor documentation.\footnote{We operationalize ``reasoning-tuned'' as endpoints whose vendor documentation describes post-training on reasoning traces or that are served with explicit thinking modes. Under this criterion GPT-5.3-Codex and Gemini 3.1 Pro qualify; Gemma 4 26B-A4B does not. The correspondence is suggestive rather than tight, and a pre-registered classification would be a stronger test.} For other model families, few-shot strategies dominate (11 of 16 models), and for GPT-5.5 and GPT-OSS-120B, zero-shot with the default system prompt is optimal.

    \item \textbf{A persistent frontier/open-source gap.} Frontier models average 75.3\% best-configuration pass rate versus 49.3\% for open-source---a 26.1\,pp gap. Only Gemma 4 31B (68.0\%) and, narrowly, GPT-OSS-120B (65.7\%) cross the frontier minimum (Claude Haiku 4.5, 60.3\%). Extended discussion in \S\ref{sec:frontier_gap}.

    \item \textbf{Prompting sensitivity is moderate across the cohort.} The largest best-vs-average configuration delta in the cohort is 5.8\,pp (GPT-OSS-20B and Gemini 3.1 Pro), and frontier models average $\overline{\Delta}$=3.4\,pp versus $\overline{\Delta}$=3.7\,pp for open-source---a much narrower spread than is sometimes reported. Prompt engineering therefore offers consistent but bounded gains across the cohort: roughly 3--6\,pp of additional pass rate is reachable by trying multiple configurations, but no model in this set transforms qualitatively under a particular prompt.
\end{itemize}

\textbf{A note on statistical significance and selection bias.} The 95\% Wilson score confidence intervals in \Cref{tab:results} quantify uncertainty over the 350-task population for each model's best configuration. Three caveats apply.

\emph{Overlapping CIs.} Several adjacent pairs in the ranking have substantially overlapping intervals---for instance, GPT-5.5 (78.9--86.7\%) and Claude Opus 4.7 (76.4--84.6\%) cannot be distinguished at the 95\% level, nor can Claude Haiku 4.5 (55.1--65.3\%) and GPT-OSS-20B (54.5--64.7\%).

\emph{Run-to-run variance.} Our single-run evaluation at temperature 0 does not capture run-to-run variance introduced by provider-side non-determinism (e.g., batching order, floating-point accumulation). Fine-grained rank differences of $<$2\,pp between adjacent models may therefore be within noise.

\emph{Best-of-7 selection bias.} Reporting the best of 7 configurations inflates the expected pass rate relative to a single pre-chosen configuration, and the inflation is larger for models with high configuration variance. In this cohort the inflation is bounded: the largest best-vs-average $\Delta$ is 5.8\,pp (GPT-OSS-20B; Gemini 3.1 Pro) and most models sit between 1 and 4\,pp, so best-of-7 selection moves point estimates by at most a few percentage points. The CIs shown are conditional on the chosen configuration and do not include this selection step. We therefore also report the Avg column in \Cref{tab:results}, which sidesteps this bias; readers interested in a selection-robust measure should weigh both columns. We recommend interpreting the results in terms of three broader tiers---top tier ($>$70\%), mid tier (55--70\%), and lower tier ($<$55\%)---which reflect statistically meaningful separations with non-overlapping CIs between tiers. Model-level rankings within a tier should be treated as approximate.

\subsection{Prompting Strategy Analysis}

\Cref{tab:prompting} shows the effect of different prompting strategies, averaged across all 16 models.

\begin{table}[h]
\caption{Effect of prompting strategies on pass rate (averaged across all 16 models). Few-shot strategies dominate on average; chain-of-thought sits between zero-shot and few-shot in aggregate but is modestly bimodal at the per-model level (see main text).}
\label{tab:prompting}
\centering
\begin{tabular}{lcccc}
\toprule
Strategy & Mean & Std & Min & Max \\
\midrule
zero-shot (default) & 55.4\% & 0.18 & 24.9\% & 83.1\% \\
zero-shot (minimal) & 54.0\% & 0.16 & 26.6\% & 78.6\% \\
zero-shot (detailed) & 50.8\% & 0.20 & 17.7\% & 80.3\% \\
\midrule
few-shot-1 & 56.7\% & 0.17 & 31.1\% & 81.4\% \\
few-shot-3 & 57.1\% & 0.16 & 32.0\% & 81.1\% \\
few-shot-5 & \textbf{57.8\%} & 0.16 & 32.3\% & 82.6\% \\
\midrule
chain-of-thought & 56.3\% & 0.17 & 31.4\% & 79.7\% \\
\bottomrule
\end{tabular}
\end{table}

\begin{figure}[h]
\centering
\includegraphics[width=\textwidth]{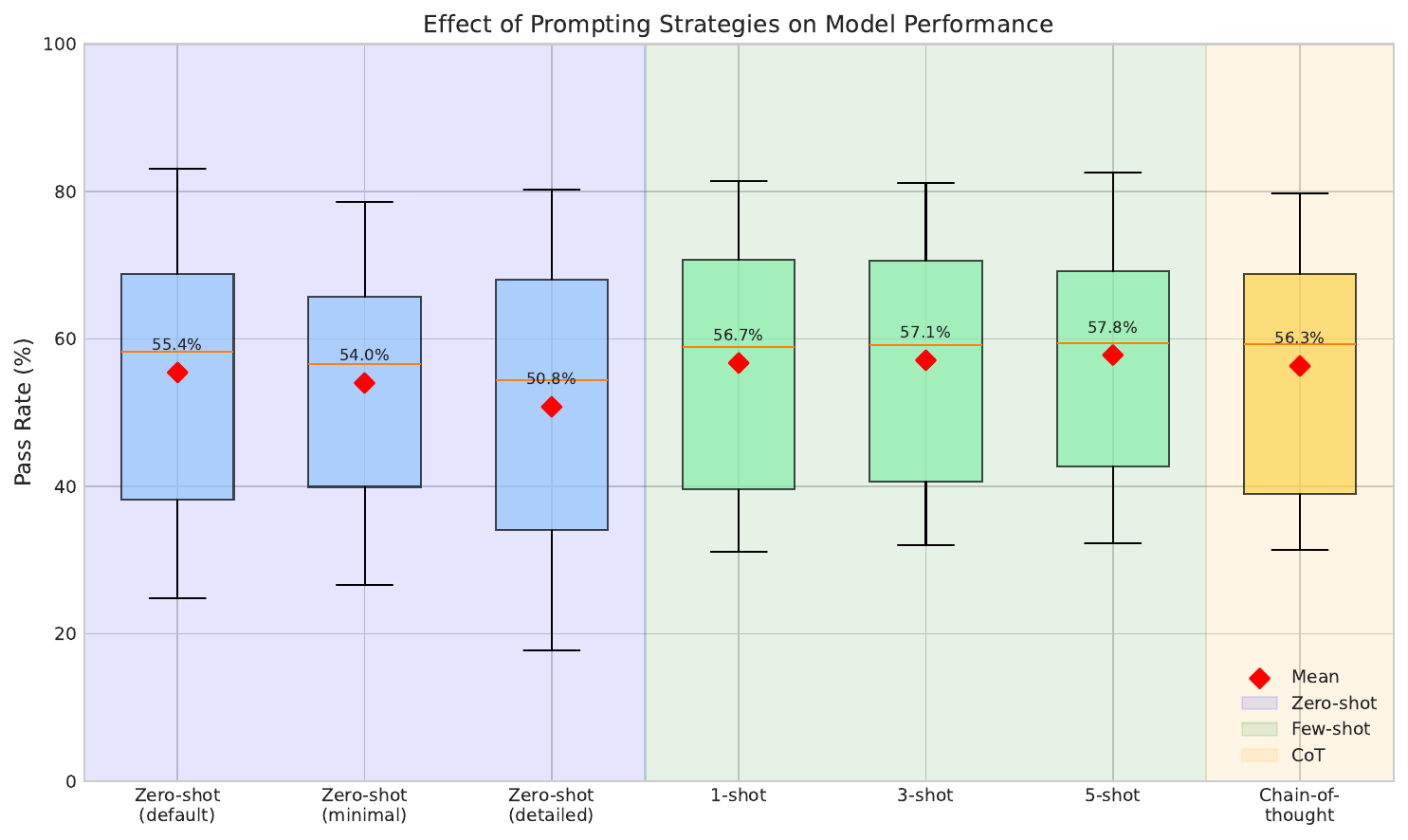}
\caption{Effect of prompting strategies on model performance. Box plots show the distribution of pass rates across all models for each strategy. Red diamonds indicate means. Few-shot prompting modestly improves the average pass rate and reduces cross-model variance.}
\label{fig:prompting}
\end{figure}

\Cref{fig:prompting} visualizes the distribution of pass rates across models for each prompting strategy.

Several patterns emerge from the prompting results.

\begin{itemize}
    \item \textbf{Few-shot is the most reliable strategy on average.} Few-shot-5 (57.8\%) is the top-scoring strategy across all 16 models, followed by few-shot-3 (57.1\%) and few-shot-1 (56.7\%). The gain over the best zero-shot variant (default, 55.4\%) is $+$2.4\,pp, and over the weakest (detailed, 50.8\%) is $+$7.0\,pp. Few-shot variants also cluster tightly (std 0.16--0.17), making per-model results more predictable.

    \item \textbf{Chain-of-thought is modestly bimodal.} In aggregate, CoT (56.3\%) sits between zero-shot default (55.4\%) and few-shot-5 (57.8\%), but this conceals a per-model split---CoT is the \emph{best} configuration for 3 of 16 models (Gemini 3.1 Pro $+$4.0\,pp, GPT-5.3-Codex $+$3.4\,pp, Gemma 4 26B-A4B $+$2.9\,pp, each relative to that model's best non-CoT configuration) and degrades performance for the remaining 13. The largest CoT penalty is Claude Sonnet 4.6 ($-$11.1\,pp), followed by GPT-OSS-20B ($-$5.4\,pp), Mistral Small 3.2 24B ($-$4.3\,pp), and GPT-5.5 ($-$3.7\,pp); the remaining models lose between 1 and 4\,pp under CoT. Full analysis in \S\ref{sec:cot_bimodal}.

    \item \textbf{System prompt wording matters.} Among zero-shot variants, the default prompt (55.4\%) outperforms minimal (54.0\%) and detailed (50.8\%) by 1.4\,pp and 4.6\,pp respectively. Detailed prompts, which prescribe specific imports and version markers, appear to conflict with conventions some models have internalized from training, forcing a choice between training priors and the prompt that is often resolved incorrectly. The minimal prompt (``Output only Python code'') provides insufficient framing for open-source models and leads to omitted imports or misread signatures. The default prompt strikes a balance between domain context and implementation freedom. The Detailed prompt also specifies ``Qiskit (version 1.0+),'' which may anchor some models toward deprecated 1.x import patterns (e.g., \texttt{qiskit.providers.aer} rather than \texttt{qiskit\_aer}) and inflate its deficit relative to a 2.x-explicit baseline. Appendix~\ref{app:prompts} presents the full text of all four system prompts.

    \item \textbf{Output length and code quality diverge.} CoT responses average 923 output tokens and 2{,}185 characters, compared with 638 tokens / 1{,}157 characters for zero-shot default and 652 tokens / 1{,}189 characters for few-shot-5. The 42\% extra tokens under CoT land mostly in natural-language reasoning rather than additional code: CoT produces 98 SyntaxErrors (versus 61 for few-shot-5) and 245 NameErrors across all models (versus 61 for few-shot-5), consistent with models drifting between a reasoning trace and the subsequent code.

    \item \textbf{Few-shot stabilizes performance.} Few-shot-3 and few-shot-5 show the lowest cross-model standard deviations (0.16), rendering performance more predictable than either CoT (0.17) or zero-shot detailed (0.20).
\end{itemize}

\begin{figure}[h]
\centering
\includegraphics[width=0.85\textwidth]{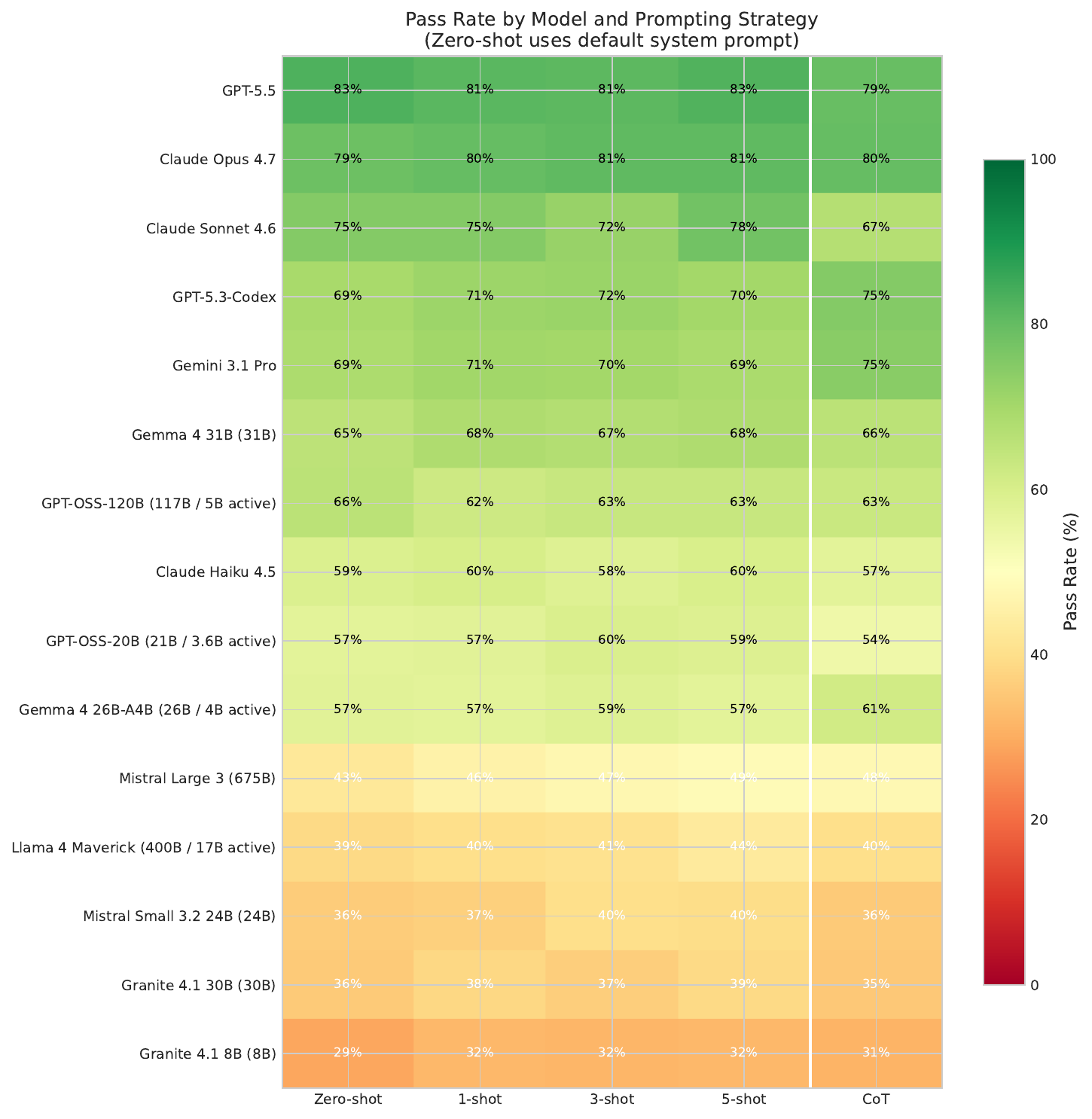}
\caption{Pass rate heatmap showing performance of the top 15 models across prompting strategies (zero-shot to 5-shot with default system prompt, plus chain-of-thought). Darker green indicates higher pass rates.}
\label{fig:heatmap}
\end{figure}

\Cref{fig:heatmap} provides a per-model breakdown. Taken together, these results suggest a simple default: few-shot-3 or few-shot-5 with the default system prompt yields the best average-case performance, with two refinements---route reasoning-tuned endpoints (in our study, GPT-5.3-Codex and Gemini 3.1 Pro) to chain-of-thought, and let the strongest instruction-followers (GPT-5.5, GPT-OSS-120B) use zero-shot, since they derive little additional benefit from in-context examples.

\subsection{Analysis by Category}

\Cref{tab:category_results} presents aggregate pass rates by category across all 16 models (using each model's best configuration).

\begin{table}[h]
\caption{Task categories ranked by aggregate pass rate across all 16 models (each using its best configuration). Total Attempts = number of tasks $\times$ 16 models, with one attempt per model per task.}
\label{tab:category_results}
\centering
\small
\begin{tabular}{lcc}
\toprule
Category & Aggregate Pass Rate & Total Attempts \\
\midrule
\multicolumn{3}{l}{\textit{Easiest categories ($>$70\%)}} \\
UnitaryPatterns & 85.4\% & 288 \\
tutorials & 82.2\% & 512 \\
SimonsAlgorithm & 82.1\% & 112 \\
BasicGates & 81.6\% & 256 \\
QEC\_BitFlipCode & 74.0\% & 192 \\
KeyDistribution\_BB84 & 70.6\% & 160 \\
TruthTables & 70.0\% & 160 \\
\midrule
\multicolumn{3}{l}{\textit{Moderate categories (50-70\%)}} \\
CHSHGame & 68.8\% & 128 \\
DeutschJozsa & 68.3\% & 240 \\
JointMeasurements & 59.6\% & 208 \\
MarkingOracles & 58.5\% & 176 \\
GraphColoring & 58.5\% & 272 \\
QFT & 57.4\% & 256 \\
Superposition & 57.1\% & 336 \\
GHZGame & 57.1\% & 112 \\
GroversAlgorithm & 55.5\% & 128 \\
PhaseEstimation & 53.6\% & 112 \\
SuperdenseCoding & 51.2\% & 80 \\
\midrule
\multicolumn{3}{l}{\textit{Hardest categories ($<$50\%)}} \\
examples & 46.9\% & 128 \\
BoundedKnapsack & 42.3\% & 272 \\
RippleCarryAdder & 42.1\% & 368 \\
Measurements & 40.3\% & 288 \\
DistinguishUnitaries & 40.0\% & 240 \\
Teleportation & 39.7\% & 224 \\
MagicSquareGame & 37.5\% & 192 \\
SolveSATWithGrover & 34.4\% & 160 \\
\bottomrule
\end{tabular}
\end{table}

\begin{figure}[h]
\centering
\includegraphics[width=\textwidth]{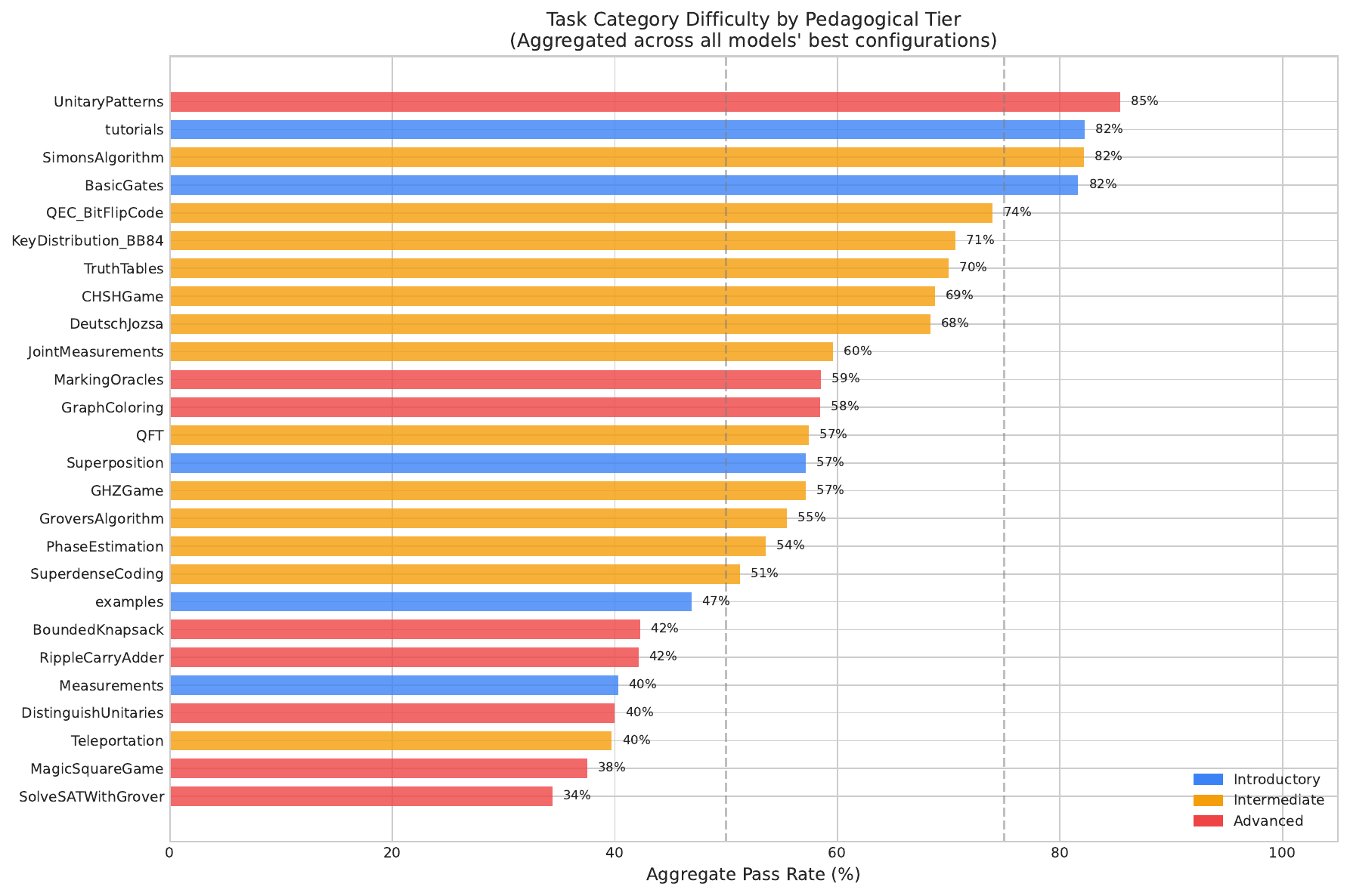}
\caption{Task category difficulty analysis. Aggregate pass rates across all models' best configurations. Categories are colored by pedagogical tier: blue (Introductory), amber (Intermediate), red (Advanced). Introductory categories generally cluster toward higher pass rates, validating the curriculum-based difficulty classification.}
\label{fig:category}
\end{figure}

\Cref{fig:category} visualizes category difficulty, with colors indicating the pedagogical tier each category belongs to.

Several patterns emerge from the category-level results.

\begin{itemize}
    \item \textbf{High-performing categories.} UnitaryPatterns (85.4\%), tutorials (82.2\%), SimonsAlgorithm (82.1\%), and BasicGates (81.6\%) are the easiest, likely reflecting simpler circuit constructions, well-documented patterns, and canonical textbook algorithms whose structure appears widely in training corpora.

    \item \textbf{Hardest category.} SolveSATWithGrover (34.4\%) combines Boolean satisfiability encoding with Grover's search---two complex components whose composition compounds difficulty. Six additional categories sit below 45\%: MagicSquareGame, Teleportation, DistinguishUnitaries, Measurements, RippleCarryAdder, and BoundedKnapsack.

    \item \textbf{Algorithm implementation versus problem encoding.} Categories requiring implementation of a known algorithm (SimonsAlgorithm 82.1\%, DeutschJozsa 68.3\%) score substantially higher than those requiring problem-to-quantum encoding (SolveSATWithGrover 34.4\%, MagicSquareGame 37.5\%, BoundedKnapsack 42.3\%). The gap is consistent with the broader observation that LLMs translate documented algorithmic structure into code more readily than they cast a classical problem into quantum primitives.

    \item \textbf{Measurement and protocol weaknesses.} Measurements (40.3\%), DistinguishUnitaries (40.0\%), and Teleportation (39.7\%) form a consistent cluster of weak spots. All three require reasoning about measurement outcomes, basis selection, or classical-communication side channels rather than pure gate construction.

    \item \textbf{Arithmetic as a distinct skill.} RippleCarryAdder (42.1\%) is among the hardest categories. Unlike pure gate tasks, it draws on classical digital-logic design (carry propagation, adder construction) applied to quantum registers and appears to require a capability that does not transfer automatically from general quantum-programming skill (cf.~the category-correlation analysis in \S\ref{sec:category_independence}).
\end{itemize}

\textbf{Normalized difficulty analysis.} Aggregate pass rates are influenced by the distribution of models evaluated. To obtain a model-independent measure, we compute \textit{normalized difficulty}: the average difference (in pp) between each model's overall pass rate and its category-specific pass rate (best configuration). Positive values indicate harder-than-average categories.

The hardest categories by this measure are SolveSATWithGrover ($+$24.7\,pp), MagicSquareGame ($+$21.5\,pp), Teleportation ($+$19.3\,pp), DistinguishUnitaries ($+$19.0\,pp), Measurements ($+$18.8\,pp), RippleCarryAdder ($+$16.9\,pp), and BoundedKnapsack ($+$16.8\,pp)---confirming these are genuinely difficult regardless of model strength. Two notable mismatches with the pedagogical tiers emerge: UnitaryPatterns ($-$26.4\,pp), classified as Advanced, is among the easiest---its tasks likely involve recognizable patterns that models handle well despite conceptual complexity. Conversely, Measurements ($+$18.8\,pp), classified as Introductory, is harder than most Intermediate and Advanced categories, indicating that measurement reasoning poses a particular challenge that pedagogical tier alone does not capture.

\textbf{Statistical caveat for small categories.} Several categories contain fewer than 10 tasks (SuperdenseCoding: 5, SimonsAlgorithm: 7, GHZGame: 7, PhaseEstimation: 7), yielding limited statistical power per model. While aggregate rates across 16 models partially mitigate this (80--112 total attempts), category-level conclusions for these groups should be interpreted with appropriate caution.

\textbf{Cross-cutting observation.} The category-level patterns above point to four drivers of difficulty---algorithm familiarity, problem-encoding load, measurement reasoning, and classical arithmetic as a separable skill---and these recur in the failure-mode breakdown: logic errors (43.0\%, \Cref{tab:errors}) far outweigh syntactic or framework errors, and align with QuanBench's \citep{quanbench2025} finding that even syntactically valid quantum circuits frequently exhibit low process fidelity. A full model-by-category breakdown is available in the supplementary materials on our GitHub repository.

\subsection{Error Analysis}

\Cref{tab:errors} presents the distribution of error types aggregated across all 16 models and 7 configurations (17,460 errors from 39,200 evaluation runs; 21,740 runs passed successfully). Each failing run produces exactly one classified error, taken from the Python exception class raised during either code extraction or test execution.

\begin{table}[h]
\caption{Error type distribution across all models and configurations (17,460 total errors from 39,200 evaluation runs). Each failing run is classified into exactly one error type. The grouped summary on the right shows logical categories for discussion.}
\label{tab:errors}
\centering
\begin{tabular}{lrrl}
\toprule
Error Type & Count & Percentage & Group \\
\midrule
AssertionError & 7,513 & 43.0\% & Logic errors (43.0\%) \\
\midrule
AttributeError & 2,018 & 11.6\% & \multirow{5}{*}{Code structure (31.6\%)} \\
ImportError & 1,685 & 9.7\% & \\
NameError & 1,152 & 6.6\% & \\
SyntaxError & 539 & 3.1\% & \\
ModuleNotFoundError & 130 & 0.7\% & \\
\midrule
MissingEntryPoint & 663 & 3.8\% & Generation failures (3.8\%) \\
\midrule
CircuitError & 1,836 & 10.5\% & \multirow{3}{*}{Qiskit API errors (13.6\%)} \\
QiskitError & 453 & 2.6\% & \\
AerError & 79 & 0.5\% & \\
\midrule
TypeError & 719 & 4.1\% & \multirow{3}{*}{Runtime/other (8.0\%)} \\
ValueError & 349 & 2.0\% & \\
Remaining ($<$1\% each) & 324 & 1.9\% & \\
\midrule
\textbf{Total} & \textbf{17,460} & \textbf{100\%} & \\
\bottomrule
\end{tabular}
\end{table}

\textbf{Error categorization.} \emph{Logic errors} (43.0\%, AssertionError) are the dominant failure mode: code runs and uses Qiskit APIs correctly but produces incorrect quantum states, indicating that quantum \textit{reasoning}---not code \textit{syntax}---is the primary limitation. \emph{Code-structure errors} (31.6\%) collect AttributeError (11.6\%), ImportError (9.7\%), NameError (6.6\%), SyntaxError (3.1\%), and ModuleNotFoundError (0.7\%); deprecated import paths (e.g., \texttt{qiskit.providers.aer} instead of \texttt{qiskit\_aer}) are a common source, and weaker open-source models (Llama 4 Scout, Granite 4.1 series) are overrepresented here. \emph{Generation failures} (3.8\%, MissingEntryPoint) occur when the model emits prose, pseudocode, or a differently-named function instead of the required entry point; in this cohort the failure mode is concentrated in open-source non-reasoning models that occasionally emit a wrongly-named function or prose-only response. \emph{Qiskit API errors} (13.6\%; CircuitError 10.5\%, QiskitError 2.6\%, AerError 0.5\%) reflect framework-specific misuse (duplicate qubit arguments, invalid circuit operations) and are slightly elevated versus earlier studies, consistent with ongoing Qiskit 2.x API evolution. \emph{Runtime/other errors} (8.0\%) comprise TypeError (4.1\%), ValueError (2.0\%), and a long tail; execution timeouts at the 30-second limit account for only 7 errors total (0.04\%), so infinite loops and intractable circuits are rare.

\begin{figure}[h]
\centering
\includegraphics[width=\textwidth]{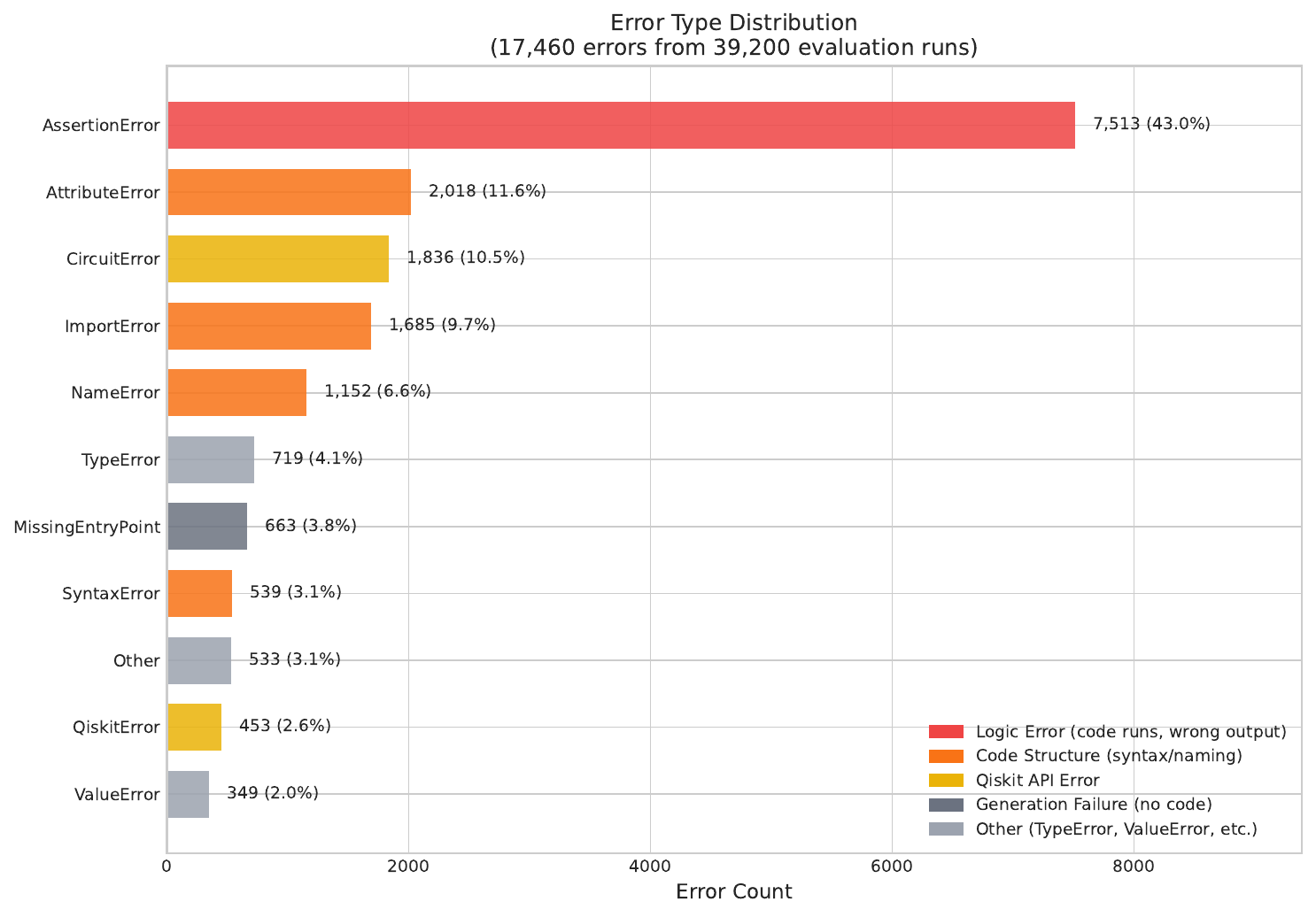}
\caption{Error type distribution across all models and configurations. Bars are colored by error category: red indicates logic errors (code runs but produces wrong output), orange indicates code structure errors (syntax/naming issues), yellow indicates Qiskit API errors, and gray indicates generation failures (no executable code produced).}
\label{fig:errors}
\end{figure}

\Cref{fig:errors} visualizes the error distribution.

Failure patterns further differ by model family. Frontier models (Claude Opus 4.7, Sonnet 4.6, Haiku 4.5; GPT-5.5, GPT-5.3-Codex; Gemini 3.1 Pro) fail primarily on AssertionError---reasoning mistakes---with comparatively low rates of syntactic or structural errors, so their dominant failure mode is logic rather than syntax. Open-source non-reasoning models (Llama 4 Scout/Maverick, Granite 4.1 8B/30B, Mistral Small 3.2 24B) show higher rates of NameError, AttributeError, and ImportError, reflecting weaker instruction following and less familiarity with Qiskit's current API surface; their SyntaxError rates are also elevated under zero-shot configurations, where the absence of in-context examples leaves more room for malformed outputs. Residual MissingEntryPoint cases in the cohort are concentrated in these same weaker open-source models, which occasionally emit a function with a wrong name or a code block missing the required entry point.

\section{Discussion}

\subsection{Dataset Characteristics and Validity}

\begin{figure}[h]
\centering
\includegraphics[width=0.85\textwidth]{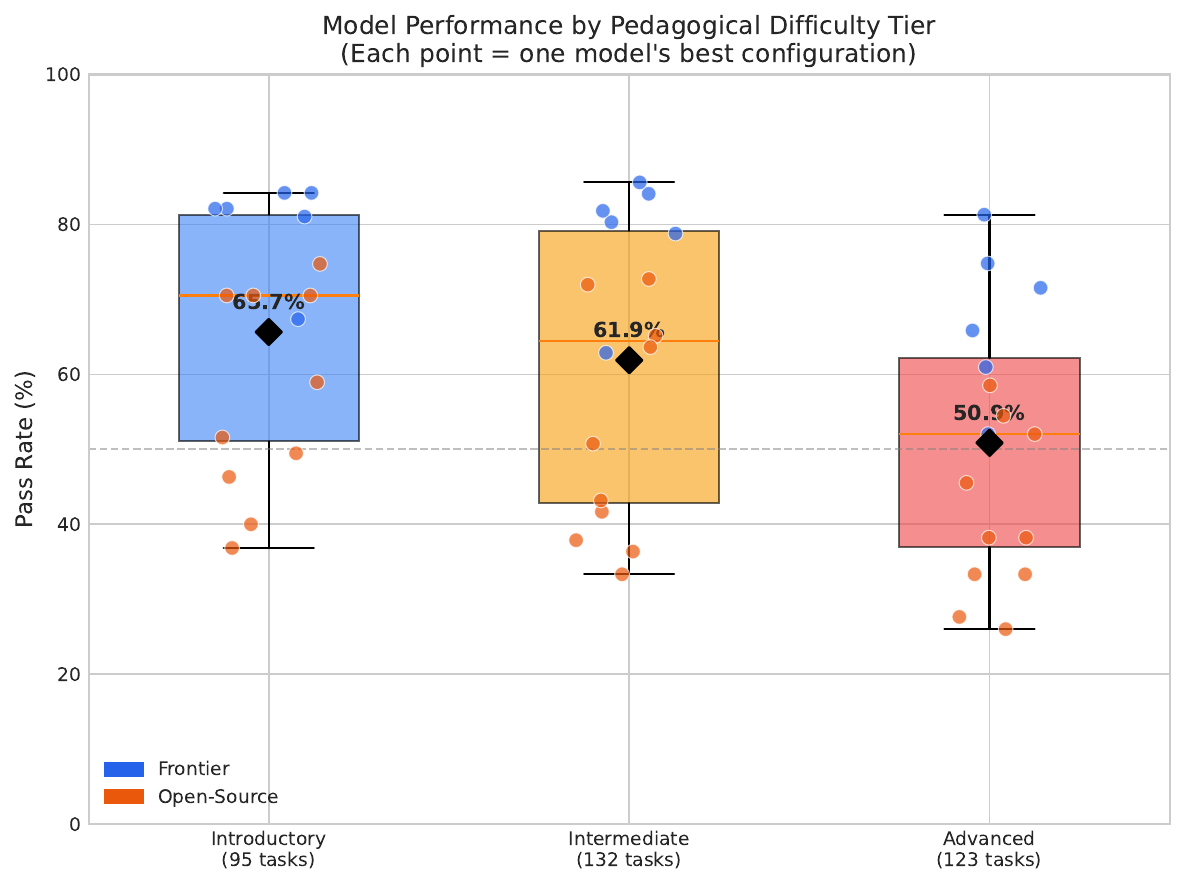}
\caption{Model performance by pedagogical difficulty tier. Box plots show the distribution of pass rates across all 16 models for each curriculum-based tier. Individual points represent models (blue = frontier, orange = open-source). Black diamonds indicate tier means. The monotonic decrease from Introductory to Advanced confirms that Microsoft's pedagogical ordering translates into measurable difficulty for LLMs.}
\label{fig:pedagogical_tiers}
\end{figure}

The evaluation results validate the benchmark along four dimensions. First, \textit{discriminative power}: best-configuration pass rates span 32.3\% to 83.1\%---a 50.8\,pp spread indicating neither ceiling nor floor effects. For comparison, QuanBench \citep{quanbench2025} reports a maximum Pass@1 of 38\% on 44 tasks, suggesting our pedagogically structured tasks produce a wider performance range. Second, \textit{category granularity}: the 26 categories reveal fine-grained capability differences that coarser benchmarks cannot provide---and, as shown in \Cref{sec:category_independence}, these categories remain sufficiently distinct to justify the 26-category granularity. Third, \textit{difficulty calibration}: \Cref{fig:pedagogical_tiers} confirms that Microsoft's pedagogical progression translates into measurable LLM difficulty, with average per-model pass rates decreasing monotonically from introductory (65.7\%) through intermediate (61.9\%) to advanced (50.9\%). Fourth, \textit{reproducibility}: deterministic verification through quantum state simulation ensures that a solution either produces the exact expected statevector or it does not---eliminating the ambiguity of approximate matching.

\subsection{Chain-of-Thought: A Bimodal Effect}
\label{sec:cot_bimodal}

Chain-of-thought prompting in this evaluation is neither uniformly helpful nor uniformly harmful. In aggregate, CoT (56.3\% mean) lies between zero-shot default (55.4\%) and few-shot-5 (57.8\%)---a narrow spread that conceals a modestly bimodal per-model effect.

Three of the sixteen models achieve their overall best pass rate under CoT: Gemini 3.1 Pro (CoT 74.6\% vs.\ 70.6\% best non-CoT; $+$4.0\,pp), GPT-5.3-Codex (CoT 75.1\% vs.\ 71.7\%; $+$3.4\,pp), and Gemma 4 26B-A4B (CoT 61.4\% vs.\ 58.6\%; $+$2.9\,pp). Two of the three (GPT-5.3-Codex and Gemini 3.1 Pro) are endpoints associated with explicit reasoning post-training, and the direction of benefit is consistent with those training regimes. For the remaining thirteen models, CoT ranks below at least one few-shot variant; the degradation is most severe for Claude Sonnet 4.6 (CoT 66.9\% vs.\ 78.0\% best non-CoT; $-$11.1\,pp), GPT-OSS-20B ($-$5.4\,pp), Mistral Small 3.2 24B ($-$4.3\,pp), GPT-5.5 ($-$3.7\,pp), and Llama 4 Maverick ($-$3.7\,pp). The remaining models lose between 1 and 4\,pp; even Claude Opus 4.7, only mildly affected overall ($-$1.1\,pp), derives no benefit from CoT.

A reasoning--code drift appears to drive most of the penalty. CoT responses average 923 output tokens against 638 for zero-shot default and 652 for few-shot-5---a 42\% increase that lands primarily in natural-language reasoning rather than additional code. Across all 16 models, CoT produces 245 NameErrors versus 61 for few-shot-5 (a 4.0$\times$ increase) and 98 SyntaxErrors versus 61, patterns consistent with models referring in later code to variables introduced only in the earlier reasoning trace, or producing malformed transitions between prose and code. A second mechanism---output-budget competition in models with internal reasoning tokens---likely matters for some endpoints not represented in this cohort; whether it explains more of the CoT penalty for thinking-tuned models in general is a question we leave to future work.

These observations have implications for prompting strategy selection. CoT appears most effective for reasoning-tuned endpoints and tends to degrade performance elsewhere; a serving setup that routes prompting by model provenance, or a training recipe that makes CoT robust across model families, would likely narrow the gap. Whether alternative CoT designs---for instance, structured decomposition or pseudocode-first reasoning---can recover the benefits observed in mathematical-reasoning benchmarks for non-reasoning models remains an open question. One caveat on effect sizes: because our few-shot baseline draws examples only from Introductory categories (\S4.3), the 57.8\% few-shot-5 mean is a conservative baseline for non-reasoning models, and same-category or difficulty-matched few-shot would likely widen, rather than close, the CoT deficit we report.

\subsection{Frontier vs. Open-Source Gap}
\label{sec:frontier_gap}

\begin{figure}[h]
\centering
\includegraphics[width=0.9\textwidth]{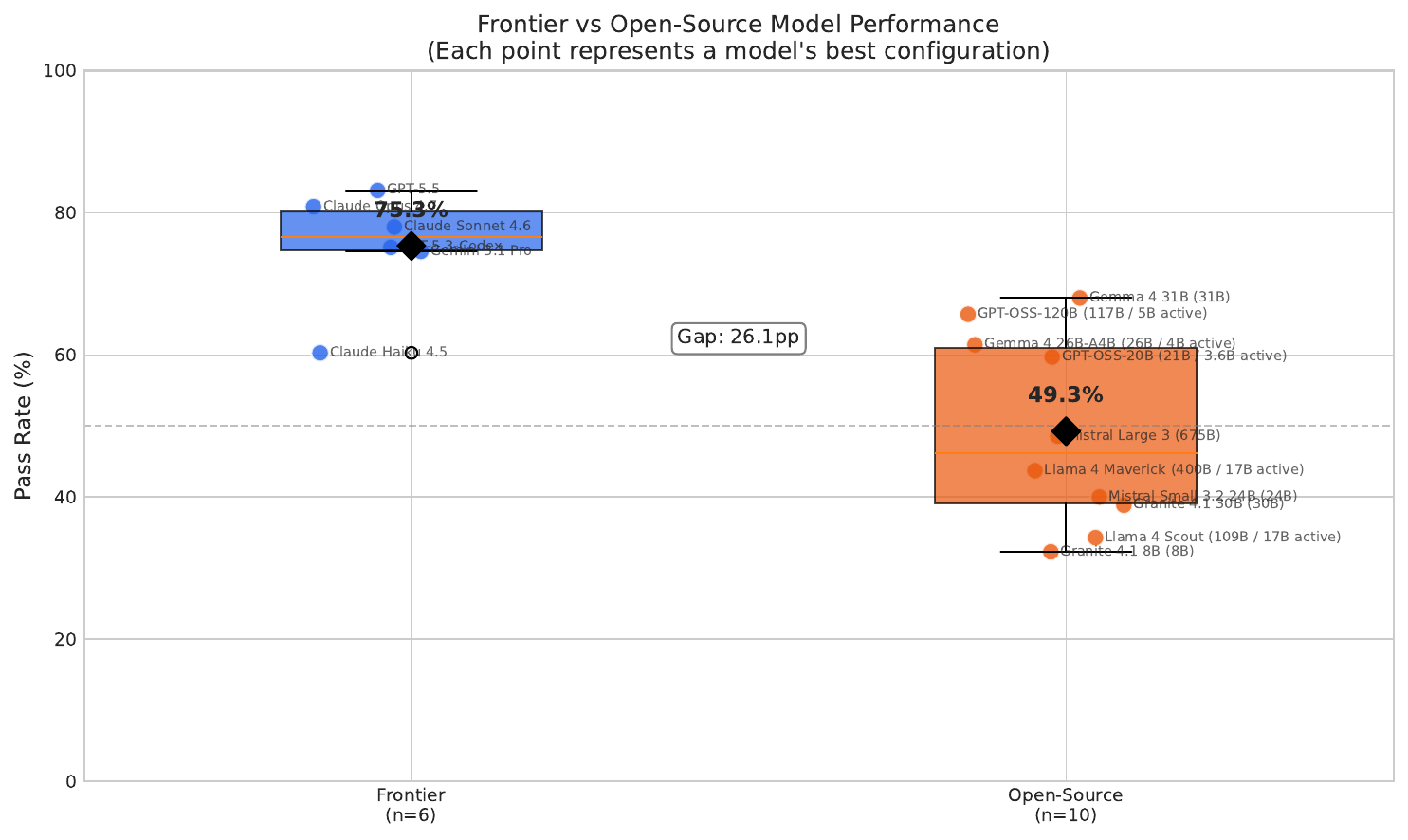}
\caption{Frontier vs. open-source model performance. Box plots show the distribution of pass rates within each category, with individual model results shown as scatter points. Black diamonds indicate category means.}
\label{fig:family}
\end{figure}

A persistent gap separates frontier and open-source models (\Cref{fig:family}). Frontier models average 75.3\% best-configuration pass rate, led by GPT-5.5 (83.1\%), Claude Opus 4.7 (80.9\%), and Claude Sonnet 4.6 (78.0\%). Open-source models average 49.3\%---a 26.1\,pp gap, consistent with prior, smaller studies. Within the open-source tier Gemma 4 31B (68.0\%) leads, followed by GPT-OSS-120B (65.7\%) and Gemma 4 26B-A4B (61.4\%); at the other end Llama 4 Scout (34.3\%) and Granite 4.1 8B (32.3\%) struggle substantially. For production quantum computing applications, frontier models remain the preferred choice; open-source alternatives are viable for exploratory or educational settings, especially the Gemma 4 family and GPT-OSS-120B. However, recent domain-specific training results---such as QUASAR's \citep{quasar2025} fine-tuned 4B model outperforming GPT-4o and GPT-5 on circuit generation---suggest this gap may be bridgeable through targeted training rather than scale alone.

\subsection{Secondary Analyses: Diversity and Category Independence}
\label{sec:solution_diversity}
\label{sec:category_independence}

Two analyses serve as construct-validity checks: that models produce genuinely diverse solutions rather than rote translations of memorized Q\# code, and that the 26 categories measure distinct capabilities rather than the same general factor.

\textbf{Solution diversity.} We computed pairwise AST similarity (Appendix~\ref{app:ast_methodology}) between the top-5 models' outputs on tasks where all five pass (217 tasks, 2{,}170 pairs; top-5 are GPT-5.5, Claude Opus 4.7, Claude Sonnet 4.6, GPT-5.3-Codex, Gemini 3.1 Pro at their best configurations). Average similarity is 0.817; 44.2\% of pairs are near-identical ($>$0.95) and 9.1\% are highly diverse ($<$0.50), with high similarity concentrated in introductory categories where one-gate solutions are effectively unique. Same-family pairs are only marginally tighter than cross-family (0.837 vs.\ 0.812 mean). This spread, combined with the dominance of logic over API-mapping errors (\Cref{tab:errors}), is more consistent with genuine per-task synthesis than with rote translation of memorized Q\# code.

\textbf{Category independence.} Pearson correlations between category-level pass rates across all 16 models (each at best configuration) are uniformly positive ($r$ from $+0.14$ to $+0.96$, mean $+0.71$), reflecting a general ``quantum programming ability'' factor whose strength varies considerably. Measurement-reasoning categories cluster tightly (Measurements $\leftrightarrow$ SuperdenseCoding $r = 0.96$, $\leftrightarrow$ DistinguishUnitaries $r = 0.94$, $\leftrightarrow$ JointMeasurements $r = 0.93$), and oracle-heavy categories form a related cluster (GraphColoring $\leftrightarrow$ MarkingOracles $r = 0.93$). At the other extreme, examples ($\bar{r} = 0.43$), MagicSquareGame ($\bar{r} = 0.47$), and GHZGame ($\bar{r} = 0.52$) draw on more idiosyncratic skills. The most representative categories---Measurements ($\bar{r} = 0.81$), and DeutschJozsa, GraphColoring, tutorials, and MarkingOracles (each $\bar{r} = 0.79$)---could serve as a compact proxy when computational budget is constrained, while the 26-category granularity remains justified for fine-grained analysis.

\subsection{Limitations}

The benchmark has several limitations.

\begin{itemize}
    \item \textbf{Simulation-based evaluation.} All tasks are verified through classical quantum simulation rather than real quantum hardware, so the benchmark does not assess capabilities related to noise mitigation, decoherence handling, or hardware-specific optimization that are critical for practical quantum computing.

    \item \textbf{Single framework.} The focus on Qiskit may not generalize to other quantum computing frameworks, and models trained primarily on alternative frameworks may be disadvantaged---although Qiskit's dominance in training data likely makes this effect small.

    \item \textbf{Educational scope.} Tasks are pedagogical rather than research-level. The benchmark does not include variational algorithms (VQE, QAOA), quantum machine learning, or modern quantum error correction schemes beyond the bit-flip code.

    \item \textbf{Translation artifacts.} The Q\#-to-Qiskit translation may have introduced subtle deviations not caught by review. Our validation confirms only that each Qiskit canonical solution passes its own translated test---partly circular, since a translation error consistent across both would slip through. We did not perform a formal Q\#-vs-Qiskit semantic-equivalence check, nor a systematic test-sensitivity audit (injecting incorrect implementations to confirm rejection); both are concrete directions for future work.

    \item \textbf{API evolution.} Qiskit is under active development. The benchmark uses Qiskit 2.x conventions (up to version 2.3), and future API changes may require dataset updates; models trained on older Qiskit documentation may also generate code using deprecated patterns such as the old \texttt{qiskit.providers.aer} import path (now \texttt{qiskit\_aer}).

    \item \textbf{Model selection.} Evaluation was restricted to models with accessible APIs at the time of benchmarking. Proprietary models may have been updated since, and some open-source models required specific hosting configurations.

    \item \textbf{Single-run evaluation.} All results are based on a single run per model per configuration at temperature 0, which minimizes but does not eliminate provider-side non-determinism (batching, floating-point ordering). Fine-grained rank differences ($<$2\,pp between adjacent models) should be interpreted with care; quantifying this variance is a concrete direction for future work (\S\ref{sec:future_directions}).

    \item \textbf{Prompting sensitivity.} Results are sensitive to system-prompt formulation, and our observation that detailed prompts underperform the default may not generalize to all models or use cases.

    \item \textbf{Potential data contamination.} The original Q\# QuantumKatas have been publicly available since 2018, so models could in principle have memorized Q\# solutions and translated them to Qiskit. Three patterns argue against this as the dominant mechanism: (i)~categories with high textbook prevalence are not uniformly easy (Measurements 40.3\% vs.\ UnitaryPatterns 85.4\%); (ii)~logic errors (43.0\%) far exceed API-mapping errors (ImportError $+$ AttributeError $+$ ModuleNotFoundError, 22.0\% combined), the opposite of what rote translation would produce; (iii)~AST similarity (\S\ref{sec:solution_diversity}) spreads from $<$0.50 to $>$0.95 with only marginal same-family clustering (0.837 vs.\ 0.812 cross-family). A pipeline-specific caveat: initial Qiskit drafts were produced with Claude Code \citep{claudecode2025} and Qiskit Code Assistant \citep{qiskitassistant2024,dupuis2024qiskit} before manual review, so Claude- and Qiskit-Assistant-lineage models share upstream tooling. Novel tasks not derived from public repositories are a natural mitigation for future work.
\end{itemize}

\section{Conclusion}

Our evaluation of 16 models across 7 prompting configurations on the Qiskit QuantumKatas benchmark yields three findings most relevant to the broader community.

First, quantum programming is within reach of current LLMs but far from solved. The best model (GPT-5.5) achieves 83.1\%, which is impressive for a specialized scientific domain---yet the hardest category (SolveSATWithGrover, 34.4\%) shows that composing multiple quantum concepts into a working solution remains a substantial challenge. The gap between algorithm implementation and problem formulation is particularly stark: models implement Simon's algorithm at 82.1\% but encode classical SAT problems into Grover's search at 34.4\%. Closing this gap will likely require advances beyond scaling alone.

Second, chain-of-thought prompting is modestly bimodal rather than uniformly beneficial or harmful. CoT is the \emph{best} configuration for three models---two of them explicitly reasoning-tuned per vendor documentation (GPT-5.3-Codex, Gemini 3.1 Pro), with Gemma 4 26B-A4B a partial exception---yet degrades performance, sometimes substantially, for the remaining thirteen. The practical implication is that prompting strategy should track model provenance: CoT for reasoning-tuned endpoints, few-shot-3 or few-shot-5 for most other models, and zero-shot only for the strongest instruction-followers (GPT-5.5, GPT-OSS-120B).

Third, the persistent 26.1 percentage-point gap between frontier and open-source models suggests that specialized scientific domains remain an area where model scale and training data quality matter considerably. Encouragingly, Gemma 4 31B (68.0\%) and GPT-OSS-120B (65.7\%) close some of this gap at a fraction of the cost, making them plausible choices for exploratory or educational settings. As open-source models continue to improve, tracking whether this gap narrows on domain-specific benchmarks like ours will be informative.

These results also have implications for quantum computing education. Frontier models are becoming viable components of automated tutoring systems, code completion tools, and solution verification---but the wide category-level variation (34.4\%--85.4\%) means such tools are broadly accurate on introductory topics while not yet trustworthy for advanced problem formulation, measurement reasoning, or quantum arithmetic.

\subsection{Future Directions}
\label{sec:future_directions}

Because this benchmark is a translation of Microsoft's QuantumKatas, its task scope is fixed by the original Q\# curriculum. We therefore distinguish two kinds of follow-up work: \textit{refinements and applications} that build directly on the released dataset, and \textit{complementary benchmarks} that this work motivates but that would require constructing new tasks beyond the QuantumKatas curriculum.

\textbf{Refinements and applications of the released benchmark.}
\begin{itemize}
    \item \textbf{Run-to-run variance quantification.} Re-running a representative subset (e.g., the top five models) over 3--5 independent samples would quantify provider-side non-determinism and indicate whether sub-2\,pp rank differences are stable.
    \item \textbf{Cross-benchmark correlation.} Evaluating a shared model set on this benchmark alongside complementary ones (Qiskit HumanEval, QuanBench, QuanBench+, QCircuitBench) would identify a compact subset that suffices for routine model comparison.
    \item \textbf{Per-task difficulty modeling.} A per-task metric using solution length, gate diversity, qubit count, and empirical pass rates would enable adaptive evaluation beyond category-level aggregation.
    \item \textbf{Multi-platform translation.} Translations of the 350 tasks to PennyLane, Cirq, or Braket---in the spirit of QuanBench+ \citep{slim2026quanbenchplus} and M2QCode \citep{m2qcode2025}---would isolate framework-specific effects from underlying quantum-programming ability.
    \item \textbf{Semantic equivalence metrics.} Replacing binary pass/fail with graded measures such as process fidelity \citep{quanbench2025} would give partial credit to circuits that are semantically close but slightly off.
    \item \textbf{Domain-specific training using this benchmark.} The 350 tasks plus their deterministic verification can serve as RL reward signals (cf.\ QUASAR \citep{quasar2025}, whose fine-tuned 4B model outperforms GPT-5 on circuit generation) or as a held-out evaluation target for instruction-tuning corpora such as QuantumLLMInstruct \citep{kashani2024quantumllminstruct}.
\end{itemize}

\textbf{Complementary benchmarks the QuantumKatas curriculum cannot cover.} The following directions go beyond Microsoft's original Q\# curriculum and are best pursued by constructing new sibling benchmarks rather than by extending this dataset:
\begin{itemize}
    \item \textbf{Noise-aware tasks.} A noise-aware sibling benchmark would assess error mitigation and noise-resilient circuit design---capabilities a simulation-based pedagogical benchmark cannot measure.
    \item \textbf{Research-level algorithms.} Variational algorithms (VQE, QAOA), quantum machine learning, and modern error-correction schemes such as surface codes go beyond textbook content and would raise the ceiling for frontier models.
    \item \textbf{Hardware-constrained optimization.} Compilation to specific qubit topologies and native gate sets would bridge textbook quantum computing and practical hardware deployment.
    \item \textbf{Contamination-resistant tasks.} Novel tasks not derived from public repositories would help disentangle memorization from genuine reasoning---a control benchmark this work motivates but cannot itself provide.
\end{itemize}

We release the benchmark dataset, evaluation framework, and all baseline results to support reproducible research on LLM capabilities in quantum computing.

\section*{Acknowledgments}

We thank Microsoft for creating the QuantumKatas and making them available as open source. The pedagogical design and comprehensive coverage of quantum computing concepts in the original Q\# implementation made this translation possible.

The translation from Q\# to Qiskit was supported by AI coding agents, including Claude Code \citep{claudecode2025} and Qiskit Code Assistant \citep{qiskitassistant2024,dupuis2024qiskit}, which assisted with API mapping, code generation, and test adaptation.

\section*{AI Writing Assistance Disclosure}

Claude Opus (Anthropic), accessed through Claude Code \citep{claudecode2025}, was used as a writing assistant during the preparation of this manuscript. Its role included drafting and revising prose, regenerating tables and figures from experimental result files, proposing structural reorganizations, and performing consistency checks across numerical claims. All scientific content, experimental design, model selection, analytical decisions, and final interpretations are the responsibility of the authors, who reviewed and edited every portion of the manuscript and figures. The benchmark dataset, evaluation framework, result files, and figure-generation scripts released alongside this paper allow readers to independently reproduce every quantitative claim in the paper.

\section*{Data Availability}

The Qiskit QuantumKatas benchmark dataset, evaluation framework, and baseline results are publicly available under the CC-BY-NC-SA-4.0 license:

\begin{itemize}
    \item \textbf{GitHub repository}: \url{https://github.com/qiskit-community/Qiskit-QuantumKatas/}
    \item \textbf{HuggingFace dataset}: \url{https://huggingface.co/datasets/Qiskit/Qiskit-QuantumKatas}
\end{itemize}

The HuggingFace dataset provides direct access to the 350 tasks in JSONL format, suitable for integration with standard ML workflows. The GitHub repository contains the full evaluation framework, prompting configurations, and scripts to reproduce all experiments reported in this paper.

\appendix

\section{Representative Tasks by Difficulty Tier}
\label{app:task_examples}

One representative task per pedagogical tier, illustrating the progression from single-gate operations to multi-component algorithm composition.

\textbf{Introductory (BasicGates/1.1 - State Flip).} A single-gate operation requiring only knowledge of the Pauli-X gate.
\begin{lstlisting}
# Input: A qubit in state |psi> = a|0> + b|1>
# Goal: Change the state to a|1> + b|0>
def state_flip(qc, q):
    qc.x(q)  # Apply Pauli-X gate
    return qc
\end{lstlisting}

\textbf{Intermediate (DeutschJozsa/1.4 - Balanced Oracle).} Requires understanding of oracles and controlled operations within a canonical quantum algorithm.
\begin{lstlisting}
# Implement oracle for f(x) = x_k (k-th bit)
def balanced_oracle(qc, x, y, k):
    qc.cx(x[k], y)  # CNOT from k-th input to output
    return qc
\end{lstlisting}

\textbf{Advanced (SolveSATWithGrover/3.1 - SAT Oracle for Grover's).} Combines Boolean satisfiability encoding with Grover's search---two complex components requiring composition of multiple quantum concepts.
\begin{lstlisting}
# Implement full Grover's search for marked state
def grovers_algorithm(qc, qubits, oracle, iterations):
    for q in qubits:
        qc.h(q)
    for _ in range(iterations):
        oracle(qc, qubits)
        for q in qubits:
            qc.h(q)
            qc.x(q)
        qc.mcp(np.pi, qubits[:-1], qubits[-1])
        for q in qubits:
            qc.x(q)
            qc.h(q)
    return qc
\end{lstlisting}

\section{Example Error Cases}
\label{app:errors}

This appendix presents representative examples of the main error types encountered during evaluation, illustrating common failure modes across models.

\subsection{Logic Error (AssertionError)}

The code executes successfully but produces incorrect quantum states. This example shows a sign error in the rotation direction:

\begin{lstlisting}[caption={Incorrect amplitude change implementation}]
# Task: BasicGates/1.4 - amplitude_change
# Expected: Apply rotation to change |0> to cos(alpha)|0> + sin(alpha)|1>

def amplitude_change(qc: QuantumCircuit, alpha: float, q: int):
    qc.ry(-2 * alpha, q)  # Wrong sign: should be +2*alpha
    return qc

# Error: AssertionError: Expected [0.866 0.5], got [0.866 -0.5]
\end{lstlisting}

The model correctly identifies that an RY gate is needed but uses the wrong rotation direction, resulting in a sign flip in the amplitude.

\subsection{API Misuse (NameError)}

The model references undefined variables or uses outdated import paths:

\begin{lstlisting}[caption={Missing import for AerSimulator}]
# Task: Measurements/1.3 - is_qubit_plus

def is_qubit_plus(qc: QuantumCircuit, q: int) -> bool:
    qc.h(q)
    qc.measure_all()
    simulator = AerSimulator()  # NameError: not imported
    job = simulator.run(qc)
    # ...

# Error: NameError: name 'AerSimulator' is not defined
# Fix: Add "from qiskit_aer import AerSimulator"
\end{lstlisting}

This error often occurs because models generate code based on older Qiskit documentation where \texttt{AerSimulator} was imported from \texttt{qiskit.providers.aer}.

\subsection{Circuit Construction Error (CircuitError)}

Qiskit-specific errors from invalid circuit operations:

\begin{lstlisting}[caption={Invalid qubit arguments to CSWAP gate}]
# Task: BasicGates/2.3 - two_qubit_gate_3

def two_qubit_gate_3(qc: QuantumCircuit, qs: list):
    qc.cswap(qs[0], qs[1], qs[0])  # Duplicate qubit!
    return qc

# Error: CircuitError: 'duplicate qubit arguments'
# Fix: Use qc.swap(qs[0], qs[1]) for simple swap
\end{lstlisting}

The model attempts to use CSWAP (Fredkin gate) but incorrectly uses the same qubit as both control and target.

\subsection{Generation Failure (MissingEntryPoint)}

Models sometimes produce a syntactically valid Python function but with the wrong name, so the test harness cannot locate the required entry point:

\begin{lstlisting}[caption={Wrong-named function defeats entry-point lookup}]
# Task: DistinguishUnitaries/1.10 - distinguish_rz_from_r1_angle
# Model: Granite 4.1 8B (few-shot-5)

# Generated code (excerpt):
def distinguish_ry_from_ry90(ry_func, qubit: int) -> int:
    """Distinguish between RY(theta) and RY..."""
    ...

# Error: Entry point 'distinguish_rz_from_r1_angle' not found
\end{lstlisting}

This occurs when the model paraphrases the problem statement into its own function name, drifts to a related-but-different task name, or omits the function definition entirely. Less commonly, weaker models exhaust their output budget mid-reasoning and emit no callable function at all.

\section{System Prompt Variants}
\label{app:prompts}

The four system prompt variants used in our evaluation (three zero-shot framings plus the chain-of-thought framing):

\textbf{Default:}
\begin{quote}
\textit{``You are an expert quantum computing programmer specializing in Qiskit.
Your task is to implement quantum computing functions using Qiskit.
Provide ONLY the Python code implementation, no explanations.
The code should be complete and ready to execute.''}
\end{quote}

\textbf{Minimal:}
\begin{quote}
\textit{``Implement the following Qiskit function. Output only Python code.''}
\end{quote}

\textbf{Detailed:}
\begin{quote}
\textit{``You are an expert quantum computing programmer with deep knowledge of Qiskit, quantum algorithms, and quantum mechanics.
Your task is to implement quantum computing functions using Qiskit (version 1.0+).
Requirements: Use standard Qiskit imports (QuantumCircuit, QuantumRegister, etc.). Implement the exact function signature provided. Return the modified QuantumCircuit. Use appropriate quantum gates from qiskit.circuit.library if needed.
Provide ONLY the Python code implementation, no explanations or markdown.''}
\end{quote}

\textbf{Chain-of-thought:}
\begin{quote}
\textit{``You are an expert quantum computing programmer specializing in Qiskit.
Your task is to implement quantum computing functions using Qiskit.
Before writing code, reason step-by-step about the quantum operations needed.
Format your response as:
THINKING: [your reasoning about the quantum circuit design]
CODE: [your Python implementation]
The code should be complete and ready to execute.''}
\end{quote}

We note that this is a single CoT prompt formulation. Alternative designs---such as structured decomposition (``first identify the required gates, then determine the qubit topology, then implement''), pseudocode-first approaches, or constraint-based reasoning---might yield different results. Our finding that CoT underperforms should therefore be interpreted as specific to this prompt style rather than a universal property of chain-of-thought reasoning for quantum tasks.

\section{AST Similarity Methodology}
\label{app:ast_methodology}

For each accepted response we extract the markdown-fenced Python code block and parse it with \texttt{ast.parse()}. We then iterate \texttt{ast.walk()} over the resulting tree to produce a linearized sequence of AST node-type names (e.g., \texttt{Module, FunctionDef, Assign, Call, Attribute, ...}). Pairwise similarity is the \texttt{difflib.SequenceMatcher} ratio over these sequences, which captures node ordering and nesting structure rather than just node-type frequencies. We compute the metric on the 217 tasks where all five top models pass at their best configuration, yielding 10 pairs per task (one per unordered model pair) and 2{,}170 pairs total. Reported summary statistics are the mean similarity, the fraction of pairs above 0.95 (near-identical) and below 0.50 (highly diverse), and same-family vs.\ cross-family means.

\bibliographystyle{unsrtnat}
\bibliography{references}

\end{document}